# Structure and Feedback in 30 Doradus II. Structure and Chemical Abundances

E. W. Pellegrini[1,2], J. A. Baldwin[2], AND G. J. Ferland[3]

[1]Department of Astronomy, University of Michigan, 500 Church Street, Ann Arbor, MI 48109, USA; pelleger@umich.edu
[2]Physics and Astronomy Department, Michigan State University, 3270 Biomedical Physical Sciences Building, East Lansing, MI 48824, USA
[3]Department of Physics and Astronomy, University of Kentucky, 177 Chemistry/Physics Building, Lexington, KY 40506, USA

## Abstract

We use our new optical-imaging and spectrophotometric survey of key diagnostic emission lines in 30 Doradus, together with CLOUDY photoionization models, to study the physical conditions and ionization mechanisms along over 4000 individual lines of sight at points spread across the face of the extended nebula, out to a projected radius 75 pc from R136 at the center of the ionizing cluster NGC 2070. We focus on the physical conditions, geometry and importance of radiation pressure on a point-by-point basis, with the aim of setting observational constraints on important feedback processes. We find that the dynamics and large scale structure of 30 Dor are set by a confined system of X-ray bubbles in rough pressure equilibrium with each other and with the confining molecular gas. Although the warm (10,000K) gas is photoionized by the massive young stars in NGC 2070, the radiation pressure does not currently play a major role in shaping the overall structure. The completeness of our survey also allows us to create a composite spectrum of 30 Doradus, simulating the observable spectrum of a spatially-unresolved, distant giant extragalactic H II region. We find that the highly simplified models used in the "strong line" abundance technique do in fact reproduce our observed lines strengths and deduced chemical abundances, in spite of the more than one order of magnitude range in the ionization parameter and density of the actual gas in 30 Dor.

*Key words:* HII regions – ISM: individual objects (30 Doradus) – ISM: structure

## 1. Introduction

Star formation is a fundamental step in the continuous chemical and structural evolution of the universe. Intense star formation is an ongoing process of cloud collapse, stellar birth and the inevitable enrichment and destruction of natal clouds. Light and winds from a newly-formed star cluster interact with the parent molecular cloud. This feedback sculpts the geometry of the regions, produces the observed emission-line spectrum, and most importantly throttles the rate of star formation.

This ongoing process is tracked in other galaxies throughout the universe via strong emission lines that are characteristic of H II regions. A significant share of our knowledge about star-formation rates, chemical abundances and abundance gradients in galaxies comes from studying emission-lines from Giant Extragalactic H II Regions (GEHRs) at intermediate to large distances (e.g. Zaritsky et al. 1994; Pettini et al. 2001; Kewley et al. 2002). Much of the current work on GEHRs depends on using the "strong-line method" (Pagel et al. 1979; Kewley & Dopita 2002;

Denicoló et al. 2002) to analyze the integrated spectra of distant, poorly-resolved cases. Because of the distances, H II regions are for the most part unresolved in galaxies beyond our own. Therefore our understanding of star formation in distant parts of the universe is largely based on models of the Interstellar Medium (ISM) and extrapolation of nearby systems to explain the observed emission lines.

To determine the impact of massive stars on further star formation, we are making detailed studies of a succession of larger and larger local star-forming regions, so far including the Orion Nebula (Pellegrini et al. 2009a; hereafter P09), M17 (Pellegrini et al. 2007), and now 30 Doradus (30 Dor). These nearby objects offer high spatial resolution of the many structural details that in more distant objects would be blended together and thus confuse the physical interpretation. The Orion Nebula is the closest. Dominated by a single O star it is the smallest scale we will consider. The O6.5 star $\theta^1$ Ori C forms a blister H II region on the surface of a background molecular cloud. Multiple studies have explored the correlation between photon flux and gas density in Orion. In P09, we showed that the detailed physical conditions of a bright ionization front (IF) called the Orion Bar were set by the absorbed energy and momentum of stellar radiation. A similar situation was found in the larger H II region M17 (Pellegrini et al. 2007) where the physical conditions were in part determined by the integrated momentum in starlight. In both regions we found that a magnetic field supports the $H^0$ region against the pressure of the integrated starlight. The relative contribution of these physical processes to the dynamics is an outstanding question in 30 Dor.

As part of the continuing development of the strong-line techniques, the process of modeling GEHRs has been refined to eliminate as many free parameters as possible. For example, Dopita et al. (2006) used the stellar synthesis code Starburst99 (Leitherer et al. 1999) to model the history of energy injection into an idealized GEHR. This provided the time-dependent evolution of a stellar population of a given metallicity, including the spectral energy distribution (SED), the total number of ionizing photons and mechanical energy released by stellar winds and supernova (SN). Other necessary input parameters included an initial mass function (IMF), and either total cluster mass for an instantaneous burst of star-formation, or a rate for continuous star formation. With the assumption of pressure balance between the H II region and surrounding material, the radius and density of the nebula at any given time are described by a shell, swept up by the mechanical energy of the SN and winds. These assumptions uniquely determine the instantaneous emission-line spectrum integrated over the volume of the idealized H II region, as a function of the chemical abundances. The observed emission-line spectrum then is used to determine the abundances.

But how well does this general treatment characterize the intrinsic properties of complex systems? The simplifying assumptions used in such studies need to be checked by careful comparison to nearby, well-resolved GEHRs where we can see in much greater detail what is actually happening.

The nearest and therefore most easily studied example of a true GEHR is 30 Doradus, sitting a mere 48.5 kpc away from us (Macri et al. 2006) in the LMC. 30 Dor has a complex and confusing geometry. As seen in Hα images such as Figure 1, taken from Paper I, its overall morphology is dominated by huge glowing arcs and cavities. These are thought to be photoionized by the star cluster NGC 2070 with the dense star clump R136 at its center (throughout the paper we will refer to R136 as our reference point on the sky because it is spatially well defined, but we will describe the ionizing radiation field as coming from the entire NGC 2070 cluster). The cavities are filled by $10^7$ K X-ray emitting gas Townsley et al. 2006). Long-slit echelle spectra (Chu & Kennicutt 1994; hereafter CK94) show that 30 Dor's velocity field includes many expanding shells on a wide range of scales. From filling-factor arguments derived from studying molecular gas, Poglitsch et al. (1995) argue that the molecular and ionized gas are in some way mixed together in a "highly fragmented" geometry. However, the overall structure of 30 Dor also has features similar to the underlying blister geometry of much smaller Galactic H II regions such as the Orion Nebula and M17. In particular, the narrow arcs that dominate H$\alpha$ images of 30 Dor (Figure 1) have molecular gas on the sides away from the central star cluster R136 (Rubio et al. 2001; M. Rubio & Probst, private communication 2009), and thus appear to trace ionization fronts seen edge-on, similar to the Bar in the Orion Nebula. How should we interpret 30 Dor? What is the physics of the feedback between stars and gas that determines what we see? Do the strong-line techniques really give the correct chemical abundances in this particular, well-resolved case?

In the first paper of this series, Pellegrini et al. (2010; hereafter Paper I) we presented a new optical imaging and spectroscopic survey of 30 Doradus. A major part of that survey was an extensive grid of long-slit optical spectra taken with the 4m Blanco telescope at CTIO, covering the key emission lines in the $\lambda\lambda$4100-7400 Å wavelength range at over 4000 points spread more-or-less uniformly over a 5×10 arcmin$^2$ area on 30 Dor. That spectroscopic grid was supplemented by additional long-slit spectra covering the $\lambda\lambda$3950-9335 Å wavelength range taken with the 4m SOAR Telescope in Chile in order to include the [S III] $\lambda$9069 line, which fell outside the limited spectral range of the CTIO data. We did not have sufficient telescope time to re-observe the entire nebula with the spatial coverage of the CTIO observations, so instead we sampled regions to the east of R136 that our CTIO data had shown to include a wide range of physical conditions. In addition, we used SOAR to obtain narrow-band direct images in [O III], H$\alpha$ and [S II]. These were used to verify the spectrophotometric calibrations across the nebula, and also provided complete high resolution spatial coverage useful for identifying discrete structures in our spatially resolved spectroscopy.

Here we use our new data set to study the processes of feedback from the massive stars, and to test the strong-line techniques. In §2 we will use a grid of Cloudy photoionization simulations to derive the physical conditions at 4238 individual points on the nebula. With these results we can test the validity of the assumption that the region can be explained as a scaled up version of smaller H II regions, in which the optical emission-line spectrum can be explained by a nebula with a constant metal and dust abundance being ionized by a single, young central stellar cluster. If 30 Dor cannot pass this test, it may be too chaotic to be adequately described by just a few global parameters, as is commonly done for distant GEHRs. In §3 we will then address the question of what has caused 30 Dor to have its present structure, and we will compare the global abundances derived from our point-by-point analysis to those that would be derived from the strong-line method.

## 2. Photoionization Simulations

## 2.1 Rationale and Purpose

As was discussed in Paper I, the observed emission line intensity ratios indicate that photoionization is the dominant excitation mechanism. Paper I also showed that there is considerable morphological evidence suggesting that the central cluster NGC 2070 is the major source of this photoionization. The emission-line intensity ratios show a general radial symmetry around a point on the sky near R136. Although the center of these circular patterns is offset to the east of R136, that can be understood as a blowout away from the observed molecular cloud adjacent to the cluster, into the lower density gas toward the east. This is very similar to what is seen on a smaller scale in M17 (Pellegrini et al. 2007). The identification of the cluster centered on R136 as the source of ionization is further supported by the lack of gas ionized locally by individual stars beyond a radius of 15 pc from R136, by the large numbers of elephant trunks and ionization fronts pointing back towards R136 from across 30 Dor, and by the thickness of edge-on ionization fronts as a function of distance from R136.

Our approach will be to compare observed diagnostic emission-line intensity ratios to the predictions of photoionization simulations using different sets of input parameters in order to determine global properties of 30 Dor. Specifically we will use observations from Paper I of H$\beta$, H$\alpha$, He I, [O I] 6300, [O II] 3725,3729, [O III] 4363, [O III] 5007, [S II] 6716,6731, [S III] 9069 and [Ar III] 7135 . Our goal is to determine the gas to dust ratio, the shape of the ionizing continuum radiation field, and the global chemical abundances averaged over the full nebula.

## 2.2 Basic Simulation Parameters

We used the plasma simulation code Cloudy (Ferland et al. 1998). All of the simulations described below begin at the illuminated face of a plane parallel cloud, externally illuminated by a source of ionizing radiation with an incident flux $Q_0/(4\pi\, r_0^2)$, where $Q_0$ is the number of H ionizing photons from the ionizing source and $r$ is the distance from the ionizing source to the cloud. In a typical H II region the fraction of neutral hydrogen is very low and the hydrogen density $n_H = n_e$. We computed a grid of photoionization simulations covering a wide range of stellar effective temperatures $T_{eff}$, ionizing photon flux $Q_0/4\pi r^2$ and $n_H$. We compare this grid to the observations from Paper I, assuming a single point-like ionization source.

For simplicity, we assumed a constant-density equation of state (EOS). A more realistic hydrostatic EOS (for example, Equation (A2) in P09) would balance the pressure gradient due to the absorption of momentum carried by ionizing starlight with the thermal gas pressure, at each point in the cloud. Using that type of EOS results in significant changes in the relative densities of the $H^+$, $H^0$ and $H_2$ regions in models that combine these three regions together in a single self–consistent simulation. But within just the $H^+$ region the gas temperature and density are roughly constant with depth, so the differences between equations of state become minimal. As a result, the predicted relative strengths of the emission lines formed in the $H^+$ zone are largely unaffected by the details of the EOS.

We have assumed a constant turbulent velocity of 4 km s$^{-1}$ to prevent unrealistic trapping of light by lines of atomic transitions.

### 2.2.1 Geometry

Each extracted spectrum contains emission from gas integrated along the line of sight through the nebula. Different lines of site intersect clouds at different angles resulting in some being viewed edge-on, and others face-on. These different geometries can alter the observed emission line strengths for the same physical conditions. Therefore how we analyze these spectra depends on how the gas is distributed and the viewing orientation.

Our grid of models includes calculations for a radiation-bounded (optically thick) geometry where all H-ionizing photons are absorbed by the gas, and also for a density-bounded (optically thin) geometry described by an optical depth $\tau_{912}$ at 912Å. In the case of the optically thick models, the physical thickness $dr$ of the $H^+$ zone will change in response to changes in density and ionizing flux wit. Generally, the ionized layer is radiation bounded, in which case the models were stopped when their temperature reached 500 K. This temperature cutoff ensures we include only regions where the observed optical lines form.

For the optically thick models, the computed emission line strengths correspond to what would be seen if the ionized cloud were viewed face-on, so that the emission coming from all depths into the $H^+$ zone was added together. This would simulate either gas seen on the ionized surfaces of a large bubble (similar to our view of the main part of the Orion Nebula, for instance), or what would be seen for a more edge-on IF with an ionized thickness small enough that the full depth of the $H^+$ zone fits into one of our one-dimensional spectral extraction windows. For our Blanco spectra the extraction window was 2.5×3.5 arcsec, corresponding to 0.6×0.8 pc. The ionized layer in gas with a typical density $n_H = n_e = 200$ cm$^{-3}$ and lying more than 50 pc (200 arcsec) from R136 would in fact fit within one of these extraction windows. This means that each extracted window can be treated independently. However, as discussed in section 2.3, a thicker layer of ionized gas viewed edge-on can affect our interpretations of models and must be dealt with carefully.

In the case of optically thin gas, the models do not reflect a particular geometry. Instead they represent a thick piece of what might be a more extensive $H^+$ region. The optically-thin gas is characterized by very weak emission from IF tracers like [N II] λ6584 and [S II] (λ6716+λ6731) relative to H recombination lines, with strengths up to an order of magnitude weaker than the approximately $0.1I(H\alpha)$ values expected for optically-thick gas. Our models assume that there is no absorbing gas between the region in question and NGC 2070. Despite the lack of total realism, the optically-thin models are still useful since the line ratios are more sensitive to the ionization parameter than to thickness.

### 2.2.2 Ionizing Stars

A reasonable estimate of the number of massive O and WR stars by spectral type in the central cluster can be made using the stellar census of the ionizing cluster by Selman et al. (1999) to characterize the properties of stars with known spectral types (Table 1). Using the conversions from spectral type to ionizing flux and effective temperature $T_{eff}$ given by Vacca et al. (1996) for the O-stars and Crowther (2007) for the WR stars, the corresponding rate of ionizing photons generated by the central cluster is $Q_0 = 7 \times 10^{51}$ s$^{-1}$, above the value $4.2 \times 10^{51}$ s$^{-1}$ estimated for this region by Crowther and Dessart (1998) for the region. This discrepancy could be resolved by using the stellar parameters given by Heap et al. (2006) which would reduce our computed $Q_0$ by about a factor of two, so we adopt $Q_0 = 4.2 \times 10^{51}$ s$^{-1}$. The benefit of fixing $Q_0$ is that we can use our models to calculate the incident ionizing flux and thus determine $r$, the distance of the gas

from NGC 2070.

Recent observations and modeling of the central star cluster have suggested that a population of extremely massive (>150M$_\odot$) stars may exist in 30 Doradus at the center of R136 (Crowther et al. 2010). These stars may account for up to 50% of the ionizing photon budget in the nebula, and could dominate the ionizing SED of the region. After accounting for the effect of winds on the emitted SED, the radiation from these stars is very similar to a 53kK O2 star (P. Crowther, private communication). We explored a small parameter space with half the ionizing flux coming from a star with $T_{eff}$ = 53kK, modeled with WMBasic (Pauldrach et al. 2001) and the other half at $T_{eff}$ = 38-41kK, but were not able to find any combination of ionization parameter and density that would explain the observed line ratios within the observed range of abundances. Models including these stars always have a ratio of [OIII] to lower ionization species outside the range of all the observations. We caution that we did not use exact models of these stars and only approximated their contribution to the ionizing continuum.

Regardless, the method we use renormalizes the theoretical SED to the total observed continuum. This accounts for all the sources without needing to know precisely how many individual stars are involved, as long as the shape of the composite continuum is correct.

### 2.2.3 X-Rays

The global X-ray emission in 30 Dor is well studied (Townsley et al. 2006) and found to show considerable surface brightness structure and to have plasma temperatures ranging from 3 – 9 x $10^6$ K. However, the X-rays are unlikely to significantly affect the ionization balance of the gas, since the measured total X-ray luminosity $L_{X-ray} = 10^{36.95}$ (Townsley et.al 2006) is 4.5 orders of magnitude lower than the UV ionizing luminosity from the stars. We have computed models with and without likely X-ray fluxes and these verify that in the H$^+$ region the X-rays have no discernable effect on the ionization structure or observed emission lines. These X-rays may be an important source of excitation in the neutral region beyond the H$^+$ gas, creating an X-ray Dissociation Region (XDR) that emits additional lines not considered here, but that is beyond the scope of this paper.

### 2.2.4 Initial Chemical Abundances

The chemical abundances in 30 Dor have been studied extensively. Empirical total gas phase abundances relative to H are usually determined by comparing the strength of recombination or collisionally excited emission lines to that of a hydrogen recombination line, using the [O III] ($\lambda$4959+ $\lambda$5007)/$\lambda$4363 ratio to estimate the gas temperature. The total abundance is measured by observing emission from all the dominant ionization states or by making an ionization completeness correction to account for unobserved ionization states. Some key studies of this type are by Vermeij & van der Hulst (2002; hereafter V02), Mathis, Chu & Peterson 1985; Rosa & Mathis 1987; Tsamis et al. 2003; and Peimbert 2003 (hereafter P03).

A more complicated, model-dependent alternative is to use photoionization codes to account for emission from all ionization states and species simultaneously. There is often degeneracy between ionizing flux, abundances and ionizing SED, but this can be broken by accounting for variable physical conditions that can affect the interpretation of observations. Previous studies of this type have been carried out by Tsamis & Pequignot (2005; hereafter TP05) and Lebouteiller

et al. (2008).

As a starting point for our simulations, we used the abundances that TP05 derived from photoionization simulations using the photoionization code NEBU (Péquignot et al. 2001). TP05 assembled a multiwavelength set of observed emission line strengths by combining published results from UV (V02), optical (P03) and IR ISO (V02) spectroscopy, and then used tailored photoionization simulations to determine the total gas-phase abundances of 30 Dor. It was concluded that, along the particular line of sight that was studied, two gas components with different chemical abundances, densities and temperatures in pressure equilibrium are required to explain the observed emission. One component represents low-temperature high-density filaments that are H and He poor, consistent with wind-blown material from pre-SN winds. We adopt here the abundances from the *other* component, which represents the homogeneous surrounding material. This is because the filaments simulated by the first component would contribute less than 10 percent of the total emission in most of the emission lines that we are studying. The abundances for the homogeneous component of TP05, their model D2, are listed in Table 2, along with other empirical and model-based abundances. The abundances quoted from P03 are those derived assuming a mean square temperature fluctuation $t^2 =0.003$. We take the scatter between these methods to represent the systematic uncertainty of using a single abundance to describe 30 Dor. The last line of Table 2 lists the abundances which we determine here, as described below.

Our initial set of simulations with Cloudy use the same abundances, stellar atmosphere and stellar parameters used by TP05 in their model D2 and vary only the incident ionizing flux and the gas density. The CoStar stellar atmosphere (Schaerer et al. 1996; Stasinska & Schaerer 1997) has an effective temperature of 38,000 K. We then modified those input parameters to arrive at a final ionizing continuum shape and chemical abundances as described below.

## 2.3 Optical Thickness and the Dust-to-Gas Ratio

In this and the following several sections we will compare the results of large grids of Cloudy models to a number of diagnostic emission-line intensity ratios measured from our spectra. These results are shown in Figures 2, 3, and 5-10, where the loci of intensity ratios measured from the thousands of Blanco points are lumped together as color-coded contour levels, while the much smaller number of SOAR spectroscopic data points are shown individually with "+" symbols. The assumption that the gas throughout 30 Dor is to first order all illuminated by the same continuum source gives these diagrams great leverage in pinning down a number of physical parameters that are assumed here to be constant throughout the nebula.

Figure 2 shows the [S II]/Hα versus [O III]/Hα diagram, which displays the tightest observed correlation between any pair of line ratios in our observed data set. This often-used diagram was first proposed by Veilleux & Osterbrock (1987) as a way of separating H II regions from active galactic nuclei. It depends on the O/S abundance ratio as well as on the gas temperature and the ionization parameter $U$ (defined in Eq. 1, below). The solid line in Figure 2 shows the series of Cloudy models computed with changing distance $r_0$ between the gas and a central ionizing source of temperature $T_{\rm eff} = 38,000$K, and using the dust-to-gas ratio $A_{\rm V}/N({\rm H}) = 1.2 \times 10^{-22}$ cm$^2$ which Weingartner and Draine (2001a, and references therein) find to give the best overall fit to the LMC as a whole. The solid line clearly does not have the same shape as the locus of the observed points. There are two alternate ways to fix this problem.

The first possible "fix" is to fit the highest ionization gas (corresponding to the lowest [S II]/Hα ratios) with density-bounded (*i.e.* optically-thin) models. Figure 3 shows the results of varying $\tau_{912}$ which causes the modeled cloud to be truncated before the formation of an IF, leading to weak [S II] emission, while having little effect on [O III]. To test for this effect, we examined our sub-arcsec-resolution SOAR [S II]/Hα images at the positions of the edge-on IFs cataloged in Table 8 of Paper I. In edge-on resolved $H^+$ layers, where by definition $\tau_{912}$ is low, we find values of log([S II]/Hα) < -1.3±0.1. We conclude that changes in $\tau_{912}$ probably do partially explain the poor fit of our optically thick, low dust models when log([S II]/Hα)≤1.4. These regions make up a very small part of the nebula, so we will simply ignore them in the main flow of our analysis.

However, this optically-thin gas cannot explain the full discrepancy at small values of the [S II]/Hα ratio, because our images also show many areas where this ratio is small but there is evidence that the gas is an optically-thick wall seen roughly face-on. The most convincing cases are several extended areas where the [S II]/Hα image is nearly featureless (so we do not appear to be looking at edge-on ionization fronts), but polycyclic aromatic hydrocarbon (PAH) emission (in Spitzer images) or CO or $H_2$ emission is seen, and/or our SOAR spectra show [O I] λ6300 emission. Such regions include over half of the area covered by the large "cavities" seen in the Hα images (Figure 1). Along these lines of sight the Hα/PAH surface brightness ratio varies by only 36 percent. Since the PAHs are photo-excited, their surface brightness is proportional to the photon flux exciting them, as is the case for Hα, implying a constant abundance of PAHs relative to H along these lines of sight. A [S II]/Hα vs. [O III]/Hβ diagram made for just these regions has points well to the left of log([S II]/Hα) = 1.4, showing that our optically thick models *do* need to be able to fit observed gas with low [S II]/Hα ratio.

The other way to lower the left-hand end of the model curve in Figure 2 is to arbitrarily decrease the dust-to-gas ratio. Our simulations of 30 Dor for the first time include a fully self-consistent treatment of gas-grain-photon interactions. Details are given in the Appendix, but in brief we use the LMC grain size distribution from Weingartner & Draine (2001b). Using the "overall" LMC dust-gas-ratio mentioned above, photoelectric heating by dust grains embedded in the ionized gas causes the thermally excited [O III] emission to exceed the observed values, creating the upturn seen in the models at small values of [S II]/Hα, which does not agree with the observations.

The dashed curve on Figure 2 shows the results of using a revised $A_V/N(H) = 0.2 \times 10^{-22}$ cm$^2$. Although the dashed curve in Figure 2 falls well below the observed data points, we will show in the next section that it can trivially be shifted upwards by a modest revision in $T_{eff}$ of the ionizing stars. It is only the shape of the curve that matters in this immediate section. Figure 4 shows that the predicted shape of the thermal spectrum in the IR is not sensitive to the actual dust abundance (provided, of course, that there is *some* dust present), and that the predicted shape for either the usual or the low dust abundance peaks at the same wavelength as the observations.

This value of $A_V/N(H)$, which is 6 times lower than the "overall" LMC ratio used above, is the highest dust-to-gas ratio which will reproduce the shape of the locus of observed data points using radiation-bounded (*i.e.* optically-thick) models. It is well below the measured values for the diffuse ISM in the LMC, which range from 0.9 to $3.4\times10^{-22}$ cm$^2$ over 17 LMC sightlines (Cartledge et al. 2005). It is also below the values measured by Dobashi et al. (2008) using stars behind molecular clouds in the LMC which shows the dust abundance in the LMC to increase towards 30 Doradus. However, these measurements did not probe the *ionized* gas in 30 Doradus, where it is possible that dust has somehow been depleted (but see the Appendix ). For a GLIMPSE sample of Milky Way HII regions photoionized by massive stars, corresponding to less extreme environments than 30 Dor, Churchwell et al (2006) found a lack of PAH emission from the central ionized regions and concluded that at least a component of the dust was being destroyed by intense UV radiation. Watson et al. (2008) studied three of these H II regions in greater detail and showed decisively that PAHs are destroyed in ionized gas, although the same data also show that some hot dust does still survive within the ionized region. Of possible relevance Yoshida et al (2011) report that dust has been separated from ionizing gas in the outflowing wind in M82, although in that case the gas has been accelerated more strongly than the dust. Still in the extreme environment of M82 or 30 Doradus, it is not totally unexpected for the ionized gas to contain less dust.

We deduce from the above points that while a small portion of our measured data are from optically thin gas, there is also a significant effect due to a rather low dust-to-gas ratio throughout the 30 Dor nebula. For the remainder of this paper, we adopt the lower dust-to-gas ratio, and fit the data with optically thick models while keeping in mind that some optically thin gas is also present in the regions of the various line-ratios diagrams corresponding to the weakest low-ionization lines.

## 2.4 Effects of the Ionizing Continuum Shape

With the dust abundance fixed, we next explored the degree to which the ionizing continuum shape is constrained by the observed emission line intensity ratios and by their variation across the nebula.

Figures 5 (a)—(d) show a set of diagnostic diagrams which compare observed emission line intensity ratios to the predicted ratios using CoStar models of different $T_{eff}$. Ideally we would like to use a CoStar atmosphere calculated at the LMC metallicity, but this is not available. However, there are only fairly small differences between the SEDs of the 40,000K CoStar models for Solar and 0.1 Solar metallicities, so we simply used a Solar metallicity atmosphere. There is a difference at the Lyman jump for the two metallicities, but since the models are normalized in $Q_0$ this is a minor detail.

Predictions and observations are shown only for cases with $100 \leq n_e \leq 200$ cm$^{-3}$. In Figures 5(a), (c) and (d), only SOAR observations are used because they include [O II] and [S III] emission lines not covered by the Blanco data. The solid lines show results for Cloudy grids with the density fixed at $10^2$ cm$^{-3}$, and an ionizing flux corresponding to distances from the central cluster between $13 \leq r \leq 140$ pc. Multiple SEDs are shown, using the CoStar continuum shapes ranging from $T_{eff}$ = 36,000 K to 44,000 K in 2,000 K steps, increasing in temperature in the direction of the arrow shown on each panel.

Figure 5a (top left) shows ([S II] λ6716+λ6731)/([S III] λ9069) vs. ([O II] λ7320+λ7330)/([O III] λ5007), which uses emission from the same ionic species as the radiation softness parameter $\eta$ originally defined by Vilchez & Pagel (1988), except that the [O II] line used here is the temperature-sensitive auroral doublet rather than λ3727. Under the assumption of an isothermal HII region it is qualitatively similar to the ratio of *S23/R23*, where *S23* and *R23* are two ratios widely used to indicate abundances of O and S.

*R23* was originally proposed by Pagel et al. (1979) and is often used in strong line abundance measurements because it includes the strongest lines from the most abundant phases of O. This diagnostic is strongly dependent on the shape of the SED and on *U,* and is only weakly dependent on abundance and density. As the effective temperature of the modeled SED increases, the emission from the low ionization species relative to those with higher ionization potential decreases. Oey et al. (2000) demonstrated that these line ratios are insensitive to the effective stellar temperature for $T_{eff} \geq 40,000$ K so we do not expect wild variations in the ratio throughout the nebula even if the ionizing SED were to change. Some of the scatter in Figure 5a may be the result of measurement errors such as imperfect night sky subtraction of the [O II] lines or improperly corrected instrumental fringing in the [S III] line, but is unlikely to indicate real variability in the shape of the ionizing radiation on pc scales. We used [O II] 7325 in our analysis even though it is sensitive to temperature changes in the nebula. The [O II] 3727 line would have been a better choice in this regard, but it was not included in our spectra because we chose our spectral coverage with the CTIO telescope in order to maintain adequate spectral resolution while including the [S II] 6716, 6731 doublet, and with SOAR in order to include [S III] 9069.

Figure 5b shows [S II]/Hα versus [O III]/Hβ as described above. Figure 5c shows our measurements of the [O I] λ6300 line, which comes from close to the $H^+$-$H^0$ boundary zone and hence offers a stringent test of how well our constant density models are fitting at the point of transition into the Photo Dissociation Region or Photon Dominated Region, known as the PDR. Figure 5d is similar to Figure 5b except that it uses only [S II]/[S III] and [O II]/[O III] line ratios, thus removing the uncertainties concerning the O/S abundance ratio. Unfortunately the auroral [O II] lines are strongly influenced by the gas temperature and so are less useful diagnostics than the more commonly-used and better-behaved [O II] 3727,3729 doublet.

From the four diagnostic diagrams in Figure 5, we found that the best-fitting models using CoStar atmospheres lie between the $T_{eff}$ = 38,000 and 39,000 K curves, so we adopt a final $T_{eff}$ = 38,500 K. The resulting grid using that temperature together with the TP05 abundances produces the best-fitting models which are shown in Figure 5 as a heavy dashed line. The fit to the locus of the observations is good except that the [O II]/[O I] ratio is under-predicted, meaning that [O I] is too strong in the models since other line ratios involving [O II] give good fits.

Since CoStar atmospheres use a simplified treatment of line blanketing which increases the predicted FUV flux, we also tried fitting a grid of WMBasic models (Pauldrach et al. 2001) of different temperatures. The WMBasic atmospheres treat the radiation transfer, including line absorption by metals, and as a result produce a significantly softer continuum shape. As a further possible continuum shape, we constructed a composite continuum made by adding together WMBasic models at different temperatures weighted by the number of stars of each spectral type as listed in Table 1. The comparison of the effects of different stellar atmospheres is a rich and complex subject (e.g. Simon-Diaz & Stasinska 2008; Stasinska & Schaerer 1997). However, a WMBasic atmosphere with $T_{eff}$ = 40,500K is very similar to the 38,500K CoStar atmosphere over the relevant wavelength range. Adopting a WMBasic atmosphere would result in a derived O/H abundance no larger than 0.1 dex higher than found below. Noting this difference, we adopt the CoStar atmosphere in order to facilitate a more direct comparison between our work and previous studies.

## 2.5 Revised Chemical Abundances

The above exercise led us to adopt a CoStar $T_{eff}$ = 38,500 K model with solar abundances as a reasonable approximation to the ionizing continuum shape produced by NGC 2070. This model gave an acceptable fit to the line ratios shown on Figure 5.

However, the resulting Cloudy models using the TP05 abundances significantly under-predict the observed electron temperatures $T_e$. This can be seen by comparing the predicted [O III] $\lambda 5007/\lambda 4363$ ratio to the observations. In Figure 6 a grid of models covering the observed range of $n(H)$ and $r_0$ are shown with $Z/Z_{TP05}$ equal to 0.00, -0.15 and -0.30 dex. This plot shows the insensitivity of this ratio to parameters other than abundance. TP05 computed the average model temperature weighted by the $H^+$ density to be $<T(n(H^+))> = 9895$ K. Using the same initial conditions and weighting our typical model produces $<T(n(H^+))> = 9310$ K. Both of these temperatures are significantly lower than our measured $F(H\beta)$-weighted mean temperature $<T_e>$ = 10,270-10,760 K.

The temperature of an H II region is largely regulated by the balance of cooling through forbidden metal lines with heating due to ionizations. The low $T_e$ of the gas could in principle be caused by our SED being too cool. However, since a harder SED is already ruled out above, we conclude that the overall metal abundances found by TP05 are too high. We cannot positively identify the source of the discrepancy between our results and those of TP05. TP05 used a complicated mixture of multiple gas components along a single line of sight in the nebula. It is possible that they accurately identified a region of higher O/H abundance, and a variation in the O/H abundance ratio could explain some of the scatter in our data, but to confirm such a result would require applying of the TP05 method to other locations in the nebula to further test it.

Oxygen lines account for about 25% of the cooling in the various models described here, therefore the single most important parameter influencing the gas temperature is the O/H abundance ratio. To arrive at a final set of models we varied O/H and the other metal abundances in lock step (leaving He unchanged) relative to the default values (i.e. with the values used in TP05 model D02). As shown in the bottom row of Table 2, the resulting O abundance log(O/H) = -3.75.

Since N, S and Ar do not dominate the cooling, their emission varies almost linearly with abundance. The abundances of these elements were adjusted so the models again matched the emission line diagnostics used in Figures 4a-d with no change in the modeled gas temperature. The resulting adopted abundances for these elements are also listed in Table 1.

This model-based method of determining the global abundances in 30 Dor is accurate for elements with well-known atomic parameters (i.e. elements from the 2nd row of the periodic table, including O and N from Table 2) and for elements where all significant ionization states are observed (O and S from Table 2). We caution that our Ar abundance estimate is based on only one emission line from a single ionization state of this 3rd row element, and therefore is quite uncertain, but no more so than any other study forced to observe a single ionization state.

We explored the effects of including the ([OII] 3727)/Hbeta ratio in the analysis by using measurements over the 3700–7150Å wavelength range at a limited number of points on the nebula, taken from Mathis et al. (1985) and Rosa & Mathis (1987). The addition of the [O II] 3727 diagnostic would lead us to adopt a higher ionizing $T_{eff}$ = 40,500K. This would not change the derived O/H abundance ratio (log(O/H) = -3.75). However it would affect the S/H ratio by +0.15dex and our N/H abundances would have to also be adjusted to log(N/H) = -5.26, bringing our N/O ratio into agreement with that found by Mathis et al. (1985). By including [OII], the ionization parameter needed to match the observations decreases. Over the major portion of the nebula this in turn decreases our derived radiation pressure, an important quantity in our study of the overall properties of 30 Dor, by about 0.3 dex. To ensure we do not underestimate the ionization parameter and $P_{stars}$, we will apply our method without the use of [O II] 3727, but noting here the consequences of not including these constraints.

Figure 7 compares the observations to the predicted intensity ratios from this final set of models that use the revised abundances. The [O III] λ5007/λ4363 ratio is in much better agreement with the observed values. Figure 8a-d is a repeat of Figure 5a-d, showing that the models with the revised abundances fit all of the other line ratios just as well as do the models with the TP05 abundances.

We find a discrepancy between the observed He I emission and the predicted value using the TP05 abundance ratio He/H = -1.10. In close agreement with V02 we find that an abundance of He/H = -1.05 is sufficient to bring the models into agreement with observations. While a difference of 0.05dex is small, the He I lines are recombination lines from $He^+$ with well-known atomic data and provide tight constraints on $\log(He^+/H^+)$. This should be nearly identical to log(He/H). The observations of Paper I do not detect any significant $He^{++}$ λ4686 nebular emission which rules out a significant fraction of $He^{++}$. There exist many arguments (Pagel et al. 1992, Osterbrock & Ferland 2006) that there should also be little to no $He^0$ coexisting with $H^+$ for ionization by hot stars. This is confirmed in all of our models, where a typical place in the nebula with $n_H = 10^2$ cm$^{-3}$ at a distance of 50 pc has an $He^0/He^+$ fraction of 0.009.

To demonstrate the robustness of the predictions of our final model, Figure 9 shows the small variation in the predicted He I λ5876/Hβ ratio for different values of $r_0$ for $\log(n_e) \leq 2.0$ (cm$^{-3}$) At $\log(n_e)$ = 1.75, log(He I 5876/Hβ) is concentrated at -0.92, consistent with an He/H abundance of -1.05 as was found by V02. Since this ratio is nearly insensitive to the $r_0$ parameter used in our models, this rules out the lower He abundance found by TP05 in favor of that found here and by V02.

Finally, Figure 10 shows the commonly used [N II]/Hα versus [O III]/Hβ diagram. Agreement between the predictions and the observations is now good in all cases. The final adopted abundances log(n(X)/n(H)) are (H:He:C:N:O:Ne:Si:S:Cl:Ar:Fe) = (0:-1.05:-4.3:-4.91:-3.75:-4.36:-5.51:-5.32:-7.16:-6.04:-5.95). Here the abundances of all elements for which we did not measure line strengths were left unchanged from the values used in TP05 model D02. An alternative strategy might have been to alter the abundances of the unobserved elements in lockstep with the O/H abundance ratio. We verified that making that change did not alter any of the predicted intensity ratios shown in Figures 6 through 9, or the values of $U$, the ionization parameter defined as the flux of photons per hydrogen atom

$$U = \frac{Q_0}{4\pi r^2 n_H} \quad (1)$$

found in the following section, by more than 2%.

## 2.6 The Physical Parameters at Each Point in the Nebula

With the SED and chemical abundances now fixed, we fit the observations to a full grid of 1254 Cloudy models with varying distance $r_0$ from R136 and density $n_H$. The [S II] λ6716/λ6731 ratio, *S2r*, is primarily dependent on the gas density and offers a strong observational constraint in this procedure. We began with a grid spacing in density such that *S2r* varied by less than 1 percent. For each point in the nebula, *S2r* was used to eliminate any model where

$$|S2r_{obs} - S2r_{model}| > 3\sigma_{obs}. \quad (2)$$

This sub-grid of $n$, $r_0$ is like the original, but with fewer possible $n_H$. For each $n_H$ in the *S2r* selected sub-sample, we compared the models with different $r_0$ by using a convergence criterion $\chi^2$ defined as

$$\chi^2(r_0, n_H) = \sum_i \left( \frac{R_{i,obs} - R_{i,model}}{\sigma_{obs} \min(R_{i,obs}, R_{i,model})} \right)^2, \quad (3)$$

where $R_{i,obs}$ and $R_{i,model}$ are the observed and modeled dereddened ratios of emission lines relative to Hβ. The emission lines in Table 3, as well as *S2r*, were used in the fitting. They were chosen because of their brightness, dependence on ionization parameter and the absence of contamination by night sky lines.

The $r_0(n)$ with the lowest $\chi^2$ was then used as the starting point for a search between neighboring models. Using linear interpolation along $r$ we found the $r_0$ which minimized $\chi^2$ for each $n$. Finally, the $n$, $r_0(n)$ pair with the lowest $\chi^2$ was selected as the best model for that particular data point. Over almost all of the nebula good fits were achieved with optically thick models.

The final result of this procedure was a grid of values of the fitted ionization parameter $U$ and electron density $n_e$ measured at every point along our slit positions. The resulting map of $U$ over the face of 30 Dor is shown in Figure 11 and discussed in Section 3.1.

## 3. Discussion

## 3.1 What is the current structure of 30 Dor?

The projected structure of 30 Dor as seen on an Hα image such as Figure 1 is dominated first of all by the very bright arcs near R136, and after that by the extensive regions of low surface brightness which have the appearance of large cavities and are often referred to by terms such as "merged SN remnants." However, despite the prominence of the bright arcs, most of the emission is from the fainter, often amorphous regions (Paper I). Here we explore the three-dimensional structure of the optically-emitting gas.

The bright arcs clearly are edge-on walls of molecular clouds whose faces are being photoionized by NGC 2070. CO maps (Poglitsch et al. 1995) show that the main bulk of the molecular gas falls in a broad N-S swathe that includes the region of the bright arcs. Direct near-infrared narrowband images (M. Rubio & R. Probst, private communication) and also long-slit infrared K-band spectra that we have obtained with the SOAR Telescope show emission in the $H_2$ 2.12 μm line coming from extensive regions on the sides of the bright arcs in the directions away from NGC 2070. In Table 8 of Paper I we cataloged a number of other examples of similar walls, but at greater distances, which also clearly are photoionized primarily by NGC 2070.

The remainder of the surface area on 30 Dor that our model-fitting identified as optically thick gas is a mix of smaller, possibly blended edge-on IFs (as is shown by the Hα/[N II] map in Figure 2 of Paper I), along with amorphous surfaces. Much of the amorphous-appearing gas is in regions where there is no clear evidence for a background molecular component in the form of CO (1-0) emission (Poglitsch et al. 1995) or clear-cut PAH emission (Meixner et al. 2006), but it still produces a fairly high [S II]/Hα ratio characteristic of an IF. If we assume that these regions are face-on IF with as-yet-undetected molecular gas behind them, this material must lie further away from us than NGC 2070 if it is indeed photoionized by direct radiation from NGC 2070.

Turning to the "cavities", Townsley et al. (2006) showed from Chandra Space Telescope maps that there is a correlation between X-ray emission and the regions of low Hα surface brightness, and argued that the correlation between these structures implies that the cavities are supported by the pressure of the hot X-ray gas. We explored the possibility that the optical emission lines in these directions were coming from optically thin inclusions of warm ($10^4$ K) gas somehow surviving within these cavities. However, we find that the optical emission from only a fairly small portion of only one of these regions (shown by the dashed oval in Figure 1) is actually fitted by optically thin models. Over most of the regions of low Hα surface brightness the optical emission lines are from optically-thick gas, which we interpret as coming from a back wall.

This interpretation is supported by the velocity structure seen in the CK94 long-slit echelle spectra of this region. Figure 3(a) of their paper shows that the bulk of the Hα emission across this "cavity" region has a velocity that appears to be somewhat redshifted, and then there is an abrupt discontinuity in the velocity structure about 230" east of R136, which is just where the eastern wall of the cavity is crossed.

These conclusions together show that, as has long been realized, the optical emission from 30 Dor traces a very layered, three-dimensional structure. We know the projection on the sky (the $x$, $y$ coordinates) of the observed features. We now use the results from our photoionization model fits to quantify as much as possible the shape and positions of these structures in the line-of-sight ($z$) direction.

The model-fitting returns a three-dimensional distance $r_{model}$ between the gas cloud in question and the ionizing source center, which we take to be R136. This distance is derived from the combination of the fitted ionization parameter $U$ and the measured gas density. We can compare it to the projected distance $r_{projected}$ to estimate $z$, the difference between the line-of-sight distance to the gas cloud and the line-of-sight distance to NGC 2070. We use

$$|z| = \left(r_{\text{model}}^2 - r_{\text{projected}}^2\right)^{1/2}. \tag{4}$$

Figure 12 shows a map of the fitted $z$ position in units of pc, at each point across the face of 30 Dor.

Slit Position 8 of Paper I passes very near to R136 in PA 0 deg, making an N-S cut through the bright arcs to the NE and SW of R136. The bright arcs are designated IFs 2 and 3 in Paper I (see Figures 2 and 14 of Paper I). The top panel of Figure 13 shows the way in which $z$ varies along Slit Position 8 as a function of the offset in pc from R136. Equation 4 does not constrain $z$ to be positive or negative, but in the vicinity of NGC 2070 we can safely assume a face-on geometry where $z$ is positive (i.e. the gas is behind the cluster). This places the background gas 40 pc behind the central cluster. The radial velocity of this region (Melnick et al. 1999) is equal to the mean radial velocity of the nebula to well within the line width of 45 km/s. The line width is similar to that of the integrated profile of the nebula (CK94) and of the faint broad component seen by (Melnick et al. 1999). An explanation put forth by Melnick et al (1999) is that this broad component is composed of discrete, unresolved high density condensations. A higher gas density would result in a lower ionization parameter making the gas appear to be farther away from NGC 2070.

The bright IFs 2 and 3, situated to either side of NGC 2070 at a projected distance of approximately 40 arcsec (9 pc), have $z \sim 0$. Thus, within the uncertainty of $Q_0$, these IFs within the bright arcs are at the same line-of-sight depth as NGC 2070.

A region north of NGC 2070 was identified by I09 to be of especially high excitation, and is also shown by our spectra to have a very high ionization parameter. This region is part of the bright arc to the NE of NGC 2070. I09 concluded that the high excitation was the result of local photoionization by a group of three WR stars which lie close to the IF at least in projection. If this were the case then the measurement of $z$ would be decreased due to enhanced flux from the WR stars. Our Slit Position 6 samples the IF that is part of the region of interest, but at a different location further from the possible influence of the WR stars mentioned by I09. At this position there are no obvious stars that might be local ionization sources. The $z$ values found along this slit position are very close to zero, similar to those found along Position 8. The simplest explanation is that despite the presence of the WR stars, NGC 2070 is still the dominant ionization source along the IF, including at the position identified by I09.

The cavity to the east of NGC 2070 is sampled by our Slit Positions 1 and 2, which again run in PA 0 deg. The middle panel of Figure 13 shows $z$ as a function of the offset along the slit in pc, with the location closest to NGC 2070 being the zero point in offset. The considerable scatter in the $z$ values can be reduced if we use an average density $n_{avg}$ in place of the separate gas densities measured at each individual point along the slit; the bottom panel in Figure 13 shows the result using $n_{avg} = 10^{1.75}$ cm$^{-3}$. In either case we find that the most of the line emission comes from gas on a back wall of the cavity that lies about 60 pc behind NGC 2070.

The above distance estimates assume that the ionized faces are always oriented to be perpendicular to the incident light from the central ionizing source. If in fact the ionization front is inclined relative to that direction by an angle $\theta$, then

$$r_{true} = r_{model}(\cos\theta)^{1/2}. \tag{5}$$

Rapid spatial changes in $\theta$ will show up on our map in Figure 12 as rapid changes in $r_{model}$. In particular, crossing the lip of an evacuated region where the gas goes from being an ionized wall with $\theta \sim 0$ to a partially-shielded region with large $\theta$ will show up as a sudden decrease in the computed $r_{model}$ as we look progressively further away from NGC 2070. This situation is sketched in Figure 14, for the region labeled "a", which is like the Bar in the Orion Nebula (see P09).

There is significant evidence that such effects exist throughout 30 Dor. Take for example the eastern limb (called IF4 in Paper I) of the large cavity to the east of NGC 2070, or the bright arc (IF3) at the front of the molecular gas cloud on the opposite side of NGC 2070. Both of these IFs border the central region of high ionization parameter. In both cases there is a strong, large-scale increase in the [S II]/H$\alpha$ ratio just beyond the peak in the [S II] emission, typical of all the other edge-on IFs, which indicates a decrease in $U$. This occurs despite a lowered gas density and similar projected distance from the central cluster. The simplest explanation for this is a lip geometry similar to that shown in example "a" of Figure 14.

An alternative explanation for a discontinuity in $\theta$ is a true geometric discontinuity in $r_{true}$, such as a free-floating filament or an optically-thick overlapping shell which blocks our line of sight to the background gas, as sketched in example "b" in Figure 14. If such an effect were important, the reddening-free 3 cm and 6 cm radio continuum images from Lazendic et al. (2003) that trace the ionized gas would show a different structure than the optical data. We have ruled out such a geometry by comparing the $A_V$ measured with optical observations to those that are made with optical-radio data.

The geometry derived here is qualitatively consistent with the velocity survey carried out by CK94), who found that 30 Dor could be separated into five distinct expanding shells (their Figure 1(c)). Shells 1, 2 and 3 mostly overlap with our survey data. These data are consistent with our interpretation of the region to the east of R136 as a large-scale bowl or shell.

Follow-up studies should bridge these two datasets and examine the observed scatter between our models and diagnostic diagrams to identify possible correlations with velocity width. This would go a step beyond the current study and examine the extent to which shock heating contributes to the large-scale emission.

We conclude that the observed H II region all across the face of 30 Dor is largely a continuous, unobstructed structure from which the large foreground cavities have been carved. It roughly takes the form of a hemispherical bowl with depth $h \sim 40$–60 pc, which we view from the open side. NGC 2070 and R136 are offset about 12 pc towards the west of the center of this bowl, and with bright arcs forming a dense inner ring with projected radius of 10–20 pc from R136 and wrapping around three sides of the star cluster.

### 3.2 What Determined the Structure of 30 Dor?

The gas component of 30 Dor has a very complex distribution in space which is seen in all wavelengths from the IR to the X-ray. The gas got where it is today in response to pressure forces that have moved it there. To try to understand better why it has taken on its current distribution, we take an inventory of the various kinds of pressure currently at work in the nebula, and of their relative strengths at different points in the gas. As a first step, we consider the thermal gas pressure observed at the IF at each point in the nebula. This is mapped in Figure 15, where the [S II] density and [O III] temperature measurements have been used to compute $P_{gas} = 2 n_e k T_e$. In Figure 16 we show the observed $n_e$, derived from our [S II] measurements. . $T_e$ should remain roughly constant with depth into an H$^+$ region. Therefore the pressure derived by combining the [S II] density with the [O III] temperature should give a reasonable estimate of the gas pressure at the IF. Most of the pressure variation is due to differences in density, as can be seen in Figure 16 which shows the $n_e$, derived from our [S II] measurements. However, there are some changes in $T_e$ across the face of 30 Dor, as can be seen in Figure 17, a map of $T_e$ as measured from the [O III] diagnostics. The unusually hot region in the lower-left corner appears to be shock heated, based on the [S II]/Hα ratio. Most of the other isolated regions with unusual temperatures correspond to the positions of bright stars which have confused our [O III] measurements. The one photoionized region that does seem to be unusually hot is at Δx ~ 170, Δy ~ -30, which has low density. Next, we want to see if $P_{gas}$ is in equilibrium with the sum of all the external pressures acting on the ionized layer.

### 3.2.1 Pressure from Stellar Radiation & Thermal gas pressure in the H$^+$ zone

Except to the east, the region of highest gas pressure and density forms a circular ring around the central cluster, reflecting the circular regions of high excitation and high density, as shown in Figures 7-12 of Paper I. Immediately outside this region (versus outside a projected radius of about 100 arcsec) the gas pressure declines by a factor of three, with the exception of the pressure in IFs 4 and 6. To determine importance of the radiation field in setting the gas pressure and density we compare $P_{gas}$ to the pressure from integrated starlight, $P_{stars}$.

To calculate the observed $P_{rad}$ we use our values of $U$. While there is significant uncertainty in the true geometry of 30 Dor, there is much less uncertainty in the ionization parameter $U$. Accounting for the inclination angle, $U$ is given by

$$U = \frac{Q_0}{4\pi r_{true}^2 c n_H} \cos\theta \tag{6}$$

or in terms of the model parameter $r_0$

$$U = \frac{Q_0}{4\pi r_0^2 c n_H}, \tag{7}$$

recovering Equation 1. Although the degeneracy between the distance to the illuminated face and the inclination angle remain unresolved, the solution for the ionization parameter is unique.

The pressure due to the momentum which the ionized gas has absorbed from the incident photons can be approximated in terms of $U$ by

$$P_{stars} = U n_H \langle h\nu \rangle \frac{L_{Bol}}{L_{ionizing}} \tag{8}$$

where $\langle h\nu \rangle$ ~ 20eV is the average energy per photon of the SED and $L_{Bol} / L_{ionizing}$ is the ratio of the bolometric luminosity to the ionizing luminosity. This approximation, which assumes that the optical depth to ionizing radiation is greater than unity, implies that 25 percent of the momentum in the radiation field will be transformed into pressure. However, much of the remaining momentum carried by the stellar radiation (*i.e.* that in the non-ionizing UV light) can still be absorbed by gas beyond the IF. For optically thick gas illuminated by the SED which we find here, 80 percent of the momentum carried by photons will be transformed into pressure on the shielded molecular gas, so we use the total luminosity $L_{Bol}$ to calculate the radiation pressure.

We next compare the observed thermal pressure to the derived radiation pressure. Figure 18 shows a map of log($P_{stars}/P_{gas}$). There is a very strong correlation between the gas pressure and the radiation pressure within the highly ionized region, indicating that the radiation pressure is having a strong effect. Here, the radiation pressure is approximately equal to the total gas pressure. Outside of the highly ionized region around NGC 2070, this ratio drops below 1/3. For example, in the outlying IF 4 and IF 6, the radiation pressure is equal to 0.25 of the gas pressure.

From this it is clear that while in the high-pressure ring the effect of direct radiation pressure is likely to be quite important, it is negligible outside of that region ( i.e. for $r_{projected}$ > 10 pc). This dichotomy can be seen in the plot of electron density $U$ in Figure 19. Two populations are evident. The first shows a correlation between $n_e$ and $U$ in the region near the cluster. This is a scaled up and more complicated form of the scaling between density and $U$ found in the Orion Nebula by Baldwin et al. (1991) and by Wen & O'Dell (1995). The second population has no correlation between density and radiation pressure suggestive that another important physical process is in equilibrium with the gas pressure, such as the X-ray emitting plasma.

### 3.2.2 Pressure from the Hot X-Ray-Emitting Gas

Since the H II region is thought to be in direct contact with the X-ray plasma, we expect the measured and modeled $P_{gas}$ to correlate with $P_{X-ray}$. In our studies of M17 and Orion (Pellegrini 2007; P09), we found that the O-stars contribute to the dynamics via stellar winds which thermalize with the existing $10^6$ - $10^7$ K plasma. This hot plasma will either escape into the ISM or be confined by the surrounding molecular cloud. In the later case, the pressure associated with this gas is thought to form the bubbles and cavities which characterize the central parts of many H II regions.

In 30 Dor, this situation is made much more complicated because of the many SNe that presumably also have contributed to the hot gas. The definitive study of the diffuse X-ray emitting structures in 30 Dor is by Townsley et al. (2006), who identified 17 unique X-ray emitting regions that lie within the area covered by our spectroscopic maps. For each region, Townsley et al. gave the temperature, the absorption-corrected luminosity, and the area on the sky. If we make the working assumption that the volume containing each region is spherical, we can compute a volume density and then a gas pressure $P_{X-ray}$ for the X-ray emitting gas in each region. $P_{X-ray}$ will be the sum of the mechanical energy input from stellar winds and SNe. $P_{X-ray}$ is shown in Figure 20 in cgs units and listed in Table 4.

Figure 20 can be divided into three regions of systematically different pressure. The region with the highest pressure is located inside the circular cavity visible in Hα to the east of NGC 2070. It is elongated to the NE and is relatively small. $P_{X-ray}$ within this bubble is approximately $10^{-9}$ dyne cm$^{-2}$. This central region is surrounded by the bright optical arcs in which the total pressure may be enhanced due to the presence of interacting stellar winds. The two other regions to the east and west of the central bubble have systematically lower pressures, $5\times10^{-10}$ and $2\times10^{-10}$ dyne cm$^{-2}$ respectively. We caution that these estimates depend on the volume according to $P \propto V^{1/2}$. Our assumptions about the geometry could be affected by a massive molecular cloud seen in $^{12}$CO maps (Poglitsch et al 1995). It is unlikely the hot X-ray emitting gas to the east of NGC 2070 is a sphere. If it were more flat, like a pancake, the line-of-sight depth would be much lower than what we have reported, so the pressures are uncertain by a factor of three or so. This means that the two lower-pressure regions mentioned above could actually be in pressure equilibrium.

The energy stored in the X-ray emitting gas is very large. The cooling time for X-ray emitting gas is approximately $t_{cool} = n_e kT / \Lambda n_e^2$. For $\Lambda = 5 \times 10^{-23}$ erg cm$^{-3}$ s$^{-1}$, the estimated emissivity at $kT = 0.5$ keV for the composition we have derived for 30 Dor ( Landi & Landini 1999 ). This cooling time is of the order $t_{cool} \sim 17$ Myr. This is longer than the oldest episode of intense star formation. As a result confined plasma will remain at X-ray emitting temperatures for the typical lifetime of a few million years for an H II region, at which time the molecular cloud will be dispersed. Therefore the heat energy currently stored in the X-ray emitting gas could have been accumulated over the lifetime of the star cluster.

The X-ray bubbles do not absorb Lyman continuum radiation. Collisionally ionized gas at a temperature 0.5 keV has a neutral H fraction of $\log(n(H^0)/n(H)) = -6.06$. We also checked to make sure that the X-ray emitting gas does not prevent ionizing radiation from reaching the outer parts of the 30 Dor nebula. Assuming a typical hydrogen density $n_H = 0.14$ cm$^{-3}$ the optical depth to ionizing radiation over the distance $r = 50$ pc to IF4 (the conspicuous edge-on IF that forms the outer wall of the large cavity to the E of R136) is $\tau = n(H_0)\ r\ \sigma = 1.1\times10^{-4}$, where $\sigma$ is the photoionization cross section of hydrogen equal to $6\times10^{-18}$ cm$^2$. Although the X-ray emitting gas has a big effect on the dynamics of the H II region, it does not impede the stars from providing the ionization.

IF 4 seems to be a clear example of a place where the X-ray pressure is dominant. The gas pressure along the wall is $5.8 \times10^{-10}$ dyne cm$^{-2}$. The expected contribution from stars in the form of radiation pressure at that distance is $2.8 \times10^{-11}$ dyne cm$^{-2}$. The other observed pressure source in the region is the diffuse X-ray emitting gas. X-ray Region 12 of Townsley et al. (2006) seems to fill the cavity that is bounded by IF4. Using the observed surface brightness and assuming a spherical geometry we estimate the electron density of the hot gas to be equal to 0.14 cm$^{-3}$. Using the temperature reported for Region 12 the thermal X-ray pressure is equal to $5\times10^{-10}$ dyne cm$^{-2}$, in close agreement with the measured pressure in the wall. In Figure 21 we show the ratio of the X-ray to gas pressure. Aside from the IFs southwest and northeast of NGC 2070 which outline the molecular cloud, the ratio of $P_{X-ray}/P_{gas}$ is between 1 and 10 for the entire region. Given the uncertainty in the assumptions regarding the geometry of the X-ray-emitting gas and how it affects $P_{X-ray}$, the typical $P_{X-ray}/P_{gas}$ ratio is higher than 1. Except in the inner regions, the pressure in the H$^+$ zone clearly is set by $P_{X-ray}$ which dominates the equation of state by an order of magnitude, and is likely driving the current outflows, expansion and compression of the remaining molecular material.

A similar overall picture is deduced from the kinematics. CK94 found that 20% to 50% of the kinetic energy throughout 30 Dor is carried by high-velocity warm (10,000K) gas in large expanding "fast shells". This is comparable to the total thermal energy ($2\times10^{51}$ ergs) in the hot X-ray emitting gas, in which these structures are located. CK94 emphasized that the boundaries of the expanding shells often do not have a one-to-one relationship to the cavity walls that are seen on Hα images such as Figure 1. Rather, the shells tend to fill parts of the cavities. In our picture, the expanding shells are likely to be the way in which SNe are heating the surrounding X-ray emitting gas, which in turn pushes up against the cavity walls. In addition, CK94 found that about half of the kinetic energy in 30 Dor is carried by 15–20 km s$^{-1}$ turbulence that pervades the whole nebula. The $\rho v^2/2$ ram pressure from these turbulent motions is about twice the thermal pressure from the warm (10,000K) gas, but it is not clear how effective such turbulence is in supporting the large-scale gas structure.

### 3.2.3 The Interplay between X-Ray and Stellar Radiation Pressures

Orion, M17 and 30 Doradus represent extremes in the physics of star-forming regions. In each case the radiation field carries momentum into the ionized gas, building in pressure to the IF. In M17, the radiation field produces a pressure an order of magnitude greater than that of the X-ray emitting gas. In 30 Dor the ionizing cluster is 50 times larger than the one in M17, yet the radiation field is mostly negligible in determining the overall dynamics of the system.

Table 5 compares the three different objects. The ratio of $L_X/L_{UV}$ is 20 and 80 times larger in 30 Dor than in M17 and Orion, respectively. This shows that the hot gas is far more important to the energetics in 30 Dor than in the other two objects. We suggest that the difference results from the long cooling time of trapped hot gas relative to the age of the systems. In the case of M17 there has been only one epoch of star formation and the X-ray gas has not built up to the level seen in 30 Doradus. Assuming that the cooling time is generally longer than any other time scale, the hot gas will reflect the integrated mechanical energy over the star formation history minus the work done driving the expansion of the region. By contrast, the impact of ionizing radiation is due only to the current episode of star formation, and decreases with the H II region diameter $D$ according to $D^{-2}$. As a result the influence of the momentum carried by radiation will generally decrease with time.

### 3.2.4 Comparison to Other Recent Studies

The work described here was carried out over the past few years (Pellegrini 2009; Pellegrini et al. 2009b) as a follow-on to our previous similar studies of M17 (Pellegrini et al. 2007) and Orion (P09). One of our basic results is that the pressure of the X-ray emitting gas is currently the major force shaping the 30 Dor nebula.

A parallel study of 30 Dor has recently been completed by Lopez et al. (2011, hereafter L11), who reached a very different conclusion. They found that radiation pressure is the dominant shaping force, exceeding the gas pressure of both the warm and hot phase out to a three-dimensional radial distance of 50 pc from R136, and stronger than the hot phase pressure (our $P_{\text{X-ray}}$) out to a three-dimensional radial distance of about 150 pc (corresponding to about 600 arcsec, *e.g.* over a region twice as large as our whole spectroscopic survey).

The key difference is that L11 assumed that the 30 Dor H II region is a fully filled sphere. The gas density at each point is derived from the radio emission measure and the radiation pressure is derived from the projected distance on the sky. We use the optical spectrum to derive the gas density, which is significantly higher because the HII region is a relatively thin layer on the surface of a larger structure. Models of the optical spectrum are used to derive the ionization parameter and resulting radiation pressure. We find lower radiation pressure because this method implicitly takes into account the difference between the projected and true separation between the star cluster and cloud.

The same assumption of a fully-filled sphere was used by L11 to derive the pressure of the hot ($10^6$ K) X-ray emitting gas, with the implication that the warm and hot gas phases somehow occupy the same volume. L11 used the same X-ray data (Townsley et al. 2006) that we used, but calculated the plasma temperature and gas density $n_X$ along each line of sight from the X-ray emission measure and again assuming a uniform density spread along the path length through the spherical volume at that point on the sky.

We used the Chandra X-ray map by Townsley et al. (2006) to determine the pressure from hot X-ray emitting gas acting at each point. As opposed to assuming a fully filled sphere (L11), we take into account the morphology observed on the Townsley et al. (2006) map that indicates that the X-ray luminosity comes from a number of cavities adding up to a smaller volume than the single large cavity that was assumed by L11. As a result of the smaller volume our density and X-ray pressure is higher, especially in region around R136.

We find that radiation pressure is currently a fairly minor perturbation over the bulk of the nebula. L11 find a peak radiation pressure, at the center of the nebula, of about $10^{-8}$ dyne cm$^{-2}$, while we find that the maximum pressure actually experienced by the $10^4$ K gas that we have studied is 30 times lower. These results are consistent with each other because of the differences in the way $P_{stars}$ is defined. This is because even at small projected radii, our derived geometry shows that there is essentially no H II gas within a true three-dimensional distance of about 10–20 pc of R136, implying that the radiation pressure from R136 and the surrounding cluster NGC 2070 has in fact pushed the gas out to form the bright arcs. The gas closest to R136, experiencing the highest radiation pressure, is found to be the dense bright arcs, but here the back-pressure $P_{gas}$ from the gas within the arcs has balanced that outwards pressure even at this comparatively small radius. The exception is the blow-out towards the east, but in that case the radiation pressure is quite small relative to the X-ray pressure by the time that the radiation reaches the far wall of the cavity at $r = 55$ pc. Although when the size of the nebula was small the early evolution of 30 Dor may have been dominated by radiation pressure, as L11 suggest, it appears that at the present time the pressure from the hot X-ray-emitting gas plays a bigger role in shaping the nebula, even if it is not fully confined.

### 3.3 The Global Abundances

The most reliable abundance measurements from this study are for He, O, N, and S relative to H. Our He abundance is consistent with all values in the literature, which is reassuring since there is very little scatter in the reported values of the He/H ratio. As is shown in Table 2, there is much less agreement for O, N and S between different studies. The reported ranges in the (O/H) abundance ratio are -3.75 ≤ log(O/H) ≤ -3.5. We find log(O/H) = -3.75, in agreement with the lower limit. Our result is driven by the need to match the observed kinetic temperature. Like O, our S abundance log(S/H) = -5.32 is within the range found in previous work, -5.32 ≤ log(S/H) ≤

-5.01. The (S/O) ratio found here, log(S/O) = -1.57, is equal to the average of the (S/O) ratios found in other studies.

The place where we find significant discrepancies with previous work is in the resulting (N/O) and (N/S) ratios (Table 2). Although our measurement of log(N/H) = -4.91 falls within the range of previous measurements, our measured log(N/O) = -1.16 and log(N/S) = 0.41 are both 0.1 in the logarithm higher than in any other study. This could conceivably be an artifact of the SED adopted here, but it is fair to say that our abundances represent a much broader average over the full 30 Dor nebula than do the previous results, and the difference may be real.

### 3.4 A Test of Strong-line Abundance Models and Methods

The technique for determining abundances employed here has not, to our knowledge, been used previously for a single H II region. It is however similar to the strong-line methods developed by Pagel et al. 1992; Tremonti et al. 2004; Kewley & Doptia 2002; Pettini & Pagel 2004 and others, which are used to study the spatially integrated spectra of distant H II regions and galaxies. However, our technique improves on the strong-line approach because it includes direct empirical measurements of $T_e$ and $n_e$. It also removes the need to describe an entire, complex nebula by a single ionization parameter and density, and instead uses the variations in ionization parameter to constrain models. Using the observed correlations between ([O III] $\lambda 5007$)/H$\beta$, ([N II] $\lambda 6584$)/H$\alpha$, ([S III] $\lambda 6312$)/([S II] $\lambda 6716 + \lambda 6731$) and [S II]/H$\alpha$ in conjunction with Cloudy models and temperature constraints, we can measure the ionization parameter and total gas phase abundances using relatively few emission lines.

The next question is: *What would the result be if 30 Dor were instead observed from a much larger distance so that only the strong-line methods could be used?* It is important to make this test because the strong-line technique is now routinely being applied to very large samples of objects seen out to large lookback times, and many big-picture results are being deduced from the resulting abundance measurements based on the integrated spectra of entire H II regions and galaxies. These include a survey of over 500 galaxies with redshifts in the range $0 < z < 1$ ( Hu et al. 2009), as well as single galaxies at $z \sim 1.7$ (Yuan & Kewley 2009) .

There are a number of warning flags that there may be systematic errors in the strong-line abundances. One is that the different strong-line methods do not agree among themselves, although calibrations can transform them all onto any one arbitrarily chosen scale (Kewley & Ellison 2008). The strong-line abundances also are known to disagree with purely empirical abundance measurements based on direct measurements of $T_e$, $n_e$ and the ionization fractions for different elements (Kennicutt et al. 2003; Tremonti et al. 2004; Kewley & Doptia 2002; Pettini & Pagel 2004). Bresolin et al. (2009) have recently determined the abundance gradient in the nearby spiral galaxy NGC 300 using empirical measurements of many separate, unresolved H II regions, and find that its slope and absolute value are in good agreement with abundances determined from stars in the same galaxy, but systematically different than the strong-line results.

Most if not all of the comparisons of strong-line to empirical techniques made to date are for entire H II regions or entire galaxies each treated as a single idealized point. For example, the Bresolin et al. (2009) study uses data for 28 H II regions spread across the face of NGC 300.

Each of those H II regions is similar in complexity and size to 30 Dor. Yin et al. (2007) sought to calibrate strong-line methods using [O III] 4363 measurements from the integrated spectra of entire galaxies.

For this reason, our study which takes into account the detailed structure within 30 Dor is a valuable complement to those earlier results for the H II regions. Here our goal is to determine how applicable the specific models used by Dopita et al. (2006) are to a complex object like 30 Dor. We will make use of all the lines observed in our survey. Many of these lines and diagnostics are not used in the popular strong line methods, but here we are developing a stricter test of the models than is offered by strong line methods alone.

We first compare our 30 Dor abundance results to those that would be determined using the strong line method of Dopita et al. (2006). These authors developed a grid of models incorporating an assumed history of stellar evolution in the H II region, which in turn determines the current SED and mechanical energy input via SNe and stellar winds. The models are parameterized by three variables which determine the physical conditions and emission line spectrum. They are the age of the cluster $t$, the metallicity $Z$, and a (gravitating mass)/pressure ratio $R$ defined as

$$R = \log_{10}\left(\frac{M_{cluster}/M_{sun}}{P/k}\right) \quad (9)$$

in cgs units.

To compare our composite spectrum to the strong-line models we will use the [S II] density and the [O III] temperature measured from our globally integrated spectrum. For the observed cluster mass of the order $10^4$-$10^5$ M$_\odot$, $n_H = 10^2$, $T$=10,500K, and $R$ is observed to be between -1 and -2. $Z$ is primarily characterized by the oxygen abundance and given in units relative to solar abundances.

In the Dopita et al. (2006) models, $Z_\odot$ is defined as log(O/H) = -3.34 . On this scale our derived oxygen abundances becomes $Z = 0.41\ Z_\odot$. The available models cover the span $-6 \leq R \leq 2$ in increments of 2. The modeled ages range from 0.1 to 4.5 Myr in increments of 0.5 Myr beginning at 0.5 Myr. Finally the available abundances are 0.05, 0.2, 0.4, 1 and 2 in units of $Z/Z_\odot$.

We compared the line strengths predicted by the Dopita et al. (2006) models to the lines measured in our 30 Dor composite spectrum, and ranked them using Equation 3 with an equal weighting for each line ratio. Given our signal-to–noise ratio, this is valid. The lines used in the comparison were He I λ6678, [O III] λ5007, [O I] λ6300,  [N II] λ6584, and ([S II] λ6716 + λ6731) relative to Hβ, together with the [S II] λ6716/ λ6731 ratio. Identifying the initial best-fit model, we interpolated the models between different values of $R$. We find the best agreement with the observations comes from models with $Z/Z_\odot$ = -0.4, $-2 < R < -1.5$ and ages $0.5 < t < 2.0$ Myr. The ages are consistent with the youngest observed population of O stars in 30 Dor. For younger ages the difference between the observed and modeled ([O III] λ5007)/ Hβ ratio is less than 10 percent, but the lower ionization lines are under-predicted by about a factor of two. The situation is reversed for older ages, when the models seem to be to be too weakly ionized, with the best fit at 2.0 Myr. We find that despite the complexities of the region, the assumptions and simplifications used in the models created by the Dopita and Kewley group are sufficient to reproduce the known physical properties of 30 Dor that we derived here using our more detailed treatment. In particular, the abundances $Z$ are in almost exact agreement.

A direct application of the strong line method (Figure 9, Kewley and Dopita 2002; hereafter KD02) yields similar results. Here we make use of the data from MCP which includes the [OII] λ3727 line necessary to calculate strong line abundances. Step one is to make an initial estimate of the O/H ratio from [N II]/([O II] 3727)= -1.35±0.02, from the point to point variations in the data. This constrains log(O/H) ≤ -3.4. R23 is observed to range from 0.88 to 0.94. Without any constraints on the ionization parameter this is consistent with log(O/H) from -3.9 to -3.5. MCP do not include the lines needed to compute S23, but we can find R23, so we estimate q= U×c, an ionization parameter from [O III]/([O II] 3727) and find q ≥ $8\times10^7$ (their figure 5). This does little to improve the uncertainty in R23, but if we iterate as KD02 suggest, the ratios of [N II]/([O II] 3727) and [N II]/[S II]= -0.18±0.014 constrain log(O/H) to lie between -3.75 and -3.6, which is consistent with all the reported abundances reported in Table 2.

A simpler abundance measurement can be made using the calibration involving only [N II]/Hα (Pettini & Pagel 2004). This is an empirical measurement based on a sample of 137 extragalactic H II regions with well-constrained (O/H) and [N II]/Hα ratios. Inserting the value of [N II]/Hα from our composite spectrum into Equation 1 from Pettini & Pagel (2004) gives a derived oxygen abundance [O/H] = -3.87 ±0.38. This is -0.12 dex lower than the value from our study or from the strong-line method of KD02) using the models from Dopita et al. (2006), but agrees with those results to within the quoted uncertainty of the method.

We have not addressed many well-known discrepancies between results from the strong-line method and those obtained from more detailed analyses. However, we have shown that in the one case of 30 Dor, the highly simplified model used by KH02 does in fact do a reasonable job of describing what is in reality a highly complex object.

## 4. Conclusions

We have combined photoionization models with empirically determined nebular temperatures and densities to measure point-by-point variations in the ionization parameter, in order to reconstruct the ionization structure of 30 Doradus. We have used optically-thick Cloudy models which include the filtering of the radiation field, and thus are capable of predicting lower-ionization lines. This brings into play species such as [O I], [N II], [S II] and [O II] along with $H^+$, [O III] and [Ar III] so that emission from all major ionization states of the entire H II region are used. We have also assumed that all gas clouds in 30 Dor see the same ionizing continuum source, and have neglected any filtering of that radiation by any intervening gas within the 30 Doradus nebula. We deduce from the derived parameters as well as from the considerable body of morphological evidence discussed in Paper I that this is a reasonable approximation in most parts of the nebula.

This technique provides us with a comprehensive measurement of the global abundances. We find an oxygen abundance 0.15 dex lower than that found in recent studies of the bright arcs, but within the observed range of reported abundances in the literature. Attempts to use the SED and abundance parameters deduced by the photoionization modeling of 30 Dor by TP05 failed. We found differences in the gas temperatures depending on the photoionization code used. This led us to adopt a new, lower oxygen abundance consistent with V02, but we find systematically higher N/O and N/S ratios than previous studies.

Despite a correlation between $n_e$ and $U$ in 30 Dor, expected if radiation pressure is important, geometric dilution over many tens of parsecs has caused this force to be relatively unimportant at the present time in most parts of the nebula. As has been suggested by other authors in previous papers, we have shown that the dynamics and large scale structure are set by a confined system of X-ray bubbles in rough pressure equilibrium with each other and with the confining molecular gas. The long cooling time of the X-ray emitting gas means that it will dominate the dynamics until it is no longer confined. This is unlike the situations in the much smaller Orion and M17 H II regions which we have also studied in detail.

We use the results from our spatially resolved survey to test the accuracy of some strong-line abundance measurement methods that are often applied to distant galaxies. Despite the general complexity and large scale changes in $U$ across 30 Dor, there is good agreement between our measured abundances and those determined using the strong-line method developed by Kewley & Dopita. Their models in which entire H II regions and H II galaxies are characterized by a single, time dependent ionization parameter accurately reproduce the global spectrum of 30 Dor. This includes correctly estimating the ratio of cluster mass to gas pressure, the cluster age and the oxygen abundance. In spite of many other warning flags (discussed above) about the accuracy of the strong-line methods, our result shows that the simplified Kewley & Dopita models do in fact do a reasonable job of describing the overall properties of an object as complex as 30 Dor.

**Acknowledgements**

We thank the referee for a number of very helpful comments. EWP gratefully acknowledges financial support from the National Science Foundation (grant AST-0305833), NASA (07-ATFP07-0124, STScI GO09736.02-A and STScI AR-10932) and Michigan State University's Center for the Study of Cosmic Evolution. JAB acknowledges support from NSF grant AST-0305833 and NASA grant NNX10AD05G. GJF gratefully acknowledges support from NASA grant 07-ATFP07-0124 and from NSF through 0607028 and 0908877.

**References**

Abel, N.P., van Hoof, P.A.M., Shaw, G., Ferland, G.J., Elwert, T., 2008, ApJ, 686, 1125

Allers, K. N., Jaffe, D. T., Lacy, J. H., Draine, B. T. & Richter, M. J. 2005, ApJ, 630, 368

Baldwin, J.A., Ferland, G.J., Martin, P. G., Corbin, M.R., Cota, S.A., Peterson, B.M. & Slettebak, A. 1991, ApJ, 374, 580 (BFM)

Bottorff, M., Lamothe, J., Momjian, E., Verner, E., Vinković, D. & Ferland, G. 1998; PASP 110, 1040

Bresolin, F., Gieren, W., Kudritzki, R., Pietrzynski, G., Urbaneja, M. A., Carroro, G., 2009, ApJ,

700, 309

Cartledeg, et al., 2005, ApJ, 630, 355

Chu, Y., and Kennicutt, R. C. 1994, ApJ, 425, 720 (CK94)

Churchwell, E., et al. 2006, ApJ, 649, 759

Crowther, P. A., 2007, ARA&A 45, 177

Crowther, P.A., Dessart, L., 1998, MNRAS, 269, 622

Crowther, P.A., Schnurr, O., Hirschi, R., Yusof, N., Parker, R.J., Goodwin, S.P. & Kassim, H.A. 2010, MNRAS, 408, 731

Crowther, P., Private Communication 2010

Denicolo, G., Terlevich, R., Terlevich, E., 2002, MNRAS, 330, 69

Dobashi, K., Bernard, J. P., Paradis, D., Reach, W. T., Kawamura, A., 2008, A&A, 484,205

Dopita, M. A., Fischera, J., Sutherland, R. S. 2006, ApJ, 167, 177

Ferland, G.J. 2001, PASP, 113, 41

Ferland, G.J., Korista, K.T., Verner, D.A., Ferguson, J.W., Kingdon, J.B., Verner, E.M., 1998, PASP, 110, 761

Garnett, D. R. 1999, IAUS, 190, 266

Heap, S.R., Lanz, T. & Hubeny, I. 2006, ApJ, 638, 409

Hu, E. M., Cowie, L. L., Kakazu, Y. Barger, A., 2009, ApJ, 698, 2014

Kennicutt, R.C., Jr., Bresolin, F., Garnett, D., 2003, ApJ, 591, 801

Kewley, L. J., Geller, M. J., Jansen, R. A. & Dopita, M. A. 2002, AJ, 124, 3135

Kewley, L.J. & Ellison, S.L. 2008, ApJ, 681, 1183

Kewley, L.J. & Dopita, M.A., 2002, ApJS, 142, 35 (KD02)

Kingdon, J.B., Ferland, G.J. & Feibelman, W.A. 1995, ApJ, 439, 793  1997, ApJ, 477, 732

Landi, E. & Landini, M., 1999, A&A, 347, 401

Lanz, T. & Hubeny, I. 2007, ApJS, 169, 83

Lazendic, J. S., Dickel, J.R. & Jones, P. A., 2003, ApJ, 569, 287

Lebouteiller, V., Bernard-Salas, J., Brandl, B., Whelan, et al., 2008, ApJ, 680, 398

Leitherer, C., et al., 1999, ApJS, 123, 3

Lopez, L., Krumholz, M.R., Bolatto, A.D., et al., 2011, ApJ, 731, 91 (L11)

Macri, M. L., Stanek, K. Z., Bersier, D., Greenhill, L. J., Reid, M. J., 2006, ApJ, 652, 1133

Mathis, J. S., Chu, Y.H. and Peterson, D. E., 1985, ApJ, 292, 155

Meixner, M., et al., 2006, AJ, 132, 2268

Melnick, J., Tenorio-Tagle, G. & Terlevich, R. 1999, MNRAS, 302, 677

Oey, M. S., Dopita, M. A., Sheilds, J. C., Smith, R. C., 2000, ApJS, 128, 511

Osterbrock, D.E. & Ferland, G. J., 2006, Astophysics of Gaseous Nebula and Active Galactic Nuclei 2nd Edition, Sausalito, California: University Science Books

Pagel, B. E. J., Edmunds, M. G., Blackwell, D. E., Chun, M. S. & Smith, G., 1979, MNRAS, 189, 95

Pagel, B. E. J., Simonson, E. A., Terlevich, R. J. & Edmonds, M., 1992, MNRAS, 255, 325

Paradis, D., et al. 2009, AJ, 138, 196

Pauldrach, A.W.A, Hoffmann, T.L., Lennon, M., 2001, A&A, 375, 161

Peimbert, A., 2003, ApJ, 584, 735 (P03)

Pellegrini, E. W., Baldwin, J. A., Ferland, G., 2010, ApJS, 191, 160   (Paper I)

Pellegrini, E.W. 2009, PhD Thesis, Michigan State University

Pellegrini, E.W., Baldwin, J.A. and Ferland, G.J. 2009b, BAAS, 41, 510

Pellegrini, E.W., Baldwin, J.A., Ferland, G.J., et al., 2007, ApJ, 658, 1119

Pellegrini, E.W., Baldwin, J.A., Ferland, G.J., Shaw, G., Heathcote, S., 2009a, ApJ, 693, 285 (P09)

Péquignot D. et al., 2001, in Ferland, G., Savin, D. W., eds, ASP Conf. Ser. Vol. 247, Spectroscopic Challenges of Photoionized Plasmas. Astron. Soc. Pac. , San Francisco , p. 533

Pettini, M., et al., 2001, ApJ, 554, 981

Pettini, M., Pagel, B., 2004, MNRAS, 348, L59

Poglitsch, A., et al., 1995, ApJ, 454, 293

Rosa, M.,  & Mathis, J. S. 1987, ApJ, 317, 163

Rubio, M., et al., 2001, ApSSS, 277, 113

Rubio, M., Private Communication 2009

Schaerer, D., de Koter, A., Schmutz, W. & Maeder, A. 1996, A&A, 310, 837

Stasinska, G. & Schaerer, D. 1997, A&A, 322, 615

Selman, F., Melnick, J., Bosch, G., Terlevich, R., 1999, A&A, 347, 532

Shaw, G., Ferland, G. J., Henney, W. J., Stancil, P. C., Abel, N. P., Pellegrini, E. W., Baldwin, J. A. & van Hoof, P. A. M. 2009, ApJ, 701, 677

Shields, G.A. 1975, ApJ, 195, 475

Simon-Diaz, S. & Stasinska, G. 2008, MNRAS, 389, 1009

Stasinska, G. & Schaerer, D. 1997, A&A, 322, 615

Tielens, A.G.G.M. 2005, The Physics and Chemistry of the Interstellar Medium, by A. G. G. M. Tielens, ISBN 0521826349. Cambridge, UK: Cambridge University Press.

Townsley, L. K., et al. 2006, AJ, 2006, 131, 2140

Tremonti, C.A., et al., 2004, ApJ, 613, 898

Tsamis, Y. G., Barlow, M. J., Liu, X. W., Danziger, I. J., Storey, P. J., 2003, MNRAS, 338, 687

Tsamis, Y. G., Pequignot, D., 2005, MNRAS, 364, 687 (TP05)

Vacca, W.D., Garmany, C.D., Shull, J., 1996, ApJ, 460, 914

van Hoof, P.A.M., Weingartner, J.C., Martin, P.G., Volk, K., Ferland, G.J, 2004, MNRAS, 350, 1330

Veilleux, S. & Osterbrock, D. E., 1987, ApJS, 63, 295

Vermeij, R. & van der Hulst, J.M. 2002, A&A, 390, 649   (V02)

Vilchez, J. M. & Pagel, B. E., 1988, MNRAS, 231, 257

Watson, C., Povich, M.S., Churchwell, E.B., et al., 2008, ApJ, 681, l341

Weingartner, J.C., Draine, B.T.  2001a, ApJ 548, 293

Weingartner, J.C., Draine, B.T., 2001b, ApJS 134, 263

Wen, Z. & O'Dell, C.R., 1995, ApJ, 438, 784

Yin, Q.Z., Liang, Y.C, Hammer, F., Brinchmann, J., et al., 2007, A&A, 462, 535

Yoshida, M., Kawabata, K., Ohyama, Y., 2011, PASJ, 63, S493

Yuan, T.T., Kewley, L.J., 2009, ApJ, 699, 161

Zaritsky, D., Kennicutt, R.C., Huchra, J.P., 1994, ApJ, 420, 87

| Table 1. The Most Massive Stars in 30 Doradus with Cataloged Spectral Type ||||
|---|---|---|---|
| Spec. Type | Number | Spec. Type | Number |
| O3Ia | 3 | O6V | 12 |
| O3III | 12 | O7V | 16 |
| O3V | 22 | O8V | 20 |
| O4V | 28 | O9 | 21 |
| O5V | 11 | WR | 19 |
| Stars within 15 arcmin cataloged in SIMBAD. ||||

| Table 2. Abundances of Selected Elements ||||||
|---|---|---|---|---|---|
| He | O | N | S | Ar | Ref. |
| --- | -3.6 | -5.1 | -5.3 | -5.8 | Garnett (1999) (average of LMC) |
| -1.05 | -3.75 | -5.42 | -5.16 | -5.86 | Vermeij et al. 2002 |
| -1.07 | -3.5 | -4.79 | -5.01 | -5.74 | P03 |
| --- | --- | --- | -5.23 | -5.68 | Lebouteiller et al. 2008 |
| --- | -3.69 | -5.21 | -5.32 | -5.84 | Mathis, Chu & Peterson (1985) |
| -1.10 | -3.6 | -4.87 | -5.19 | -5.89 | TP05 |
| -1.05 | -3.75 | -4.91 | -5.32 | -5.99 | This work |
| Units of log[N(X)/N(H)] ||||||

| Table 3. Emission Lines Used In Model Fitting ||
|---|---|
| Species | $\lambda_{rest}$ (Å) |
| H I | 4861 |
| [O III] | 5007 |
| He I | 5875 |
| [N II] | 6584 |
| [S II] | 6717 |
| [S II] | 6731 |
| He I | 7065 |
| [Ar III] | 7135 |

| Table 4. X-ray Pressure of Selected Regions in 30 Doradus ||
|---|---|
| Region Number Townsley et al. 2006 | Pressure[1] |
| 1 | 1.7 |
| 2 | 1.3 |
| 3 | 3.0 |
| 4 | 10.2 |
| 5 | 7.9 |
| 6 | 10.0 |
| 7 | 13.4 |
| 8 | 7.9 |
| 9 | 7.2 |
| 10 | 5.8 |
| 12 | 5.0 |
| 13 | 6.0 |
| 14 | 6.6 |
| 15 | 5.6 |
| 16 | 2.8 |
| 17 | 2.4 |
| 20 | 3.0 |
| 22 | 1.5 |
| 24 | 1.4 |

[1] Units of $10^{-10}$ dyne cm$^{-2}$

| Table 5. H II Region Energetics ||||
|---|---|---|---|
| Parameter | 30 Doradus | M17 | Orion |
| $Q$(H) s$^{-1}$ | $5.0 \times 10^{51}$ | $1.4 \times 10^{50}$ | $1.0 \times 10^{49}$ |
| $L$(UV) erg s$^{-1}$ | $3.2 \times 10^{41}$ | $4.7 \times 10^{39}$ | $3.0 \times 10^{38}$ |

| $L_\text{X-ray}$ (diffuse) erg s$^{-1}$ | $4.6\times10^{36}$ | $3.4\times10^{33}$ | $5.5\times10^{31}$ |
| --- | --- | --- | --- |
| $L_\text{X-ray}/L(\text{UV})$ | $1.4\times10^{-5}$ | $7.2\times10^{-7}$ | $1.8\times10^{-7}$ |
| Cooling Time $10^6$ yr | 17.0 | 7.0 | 1.8-3.9 |

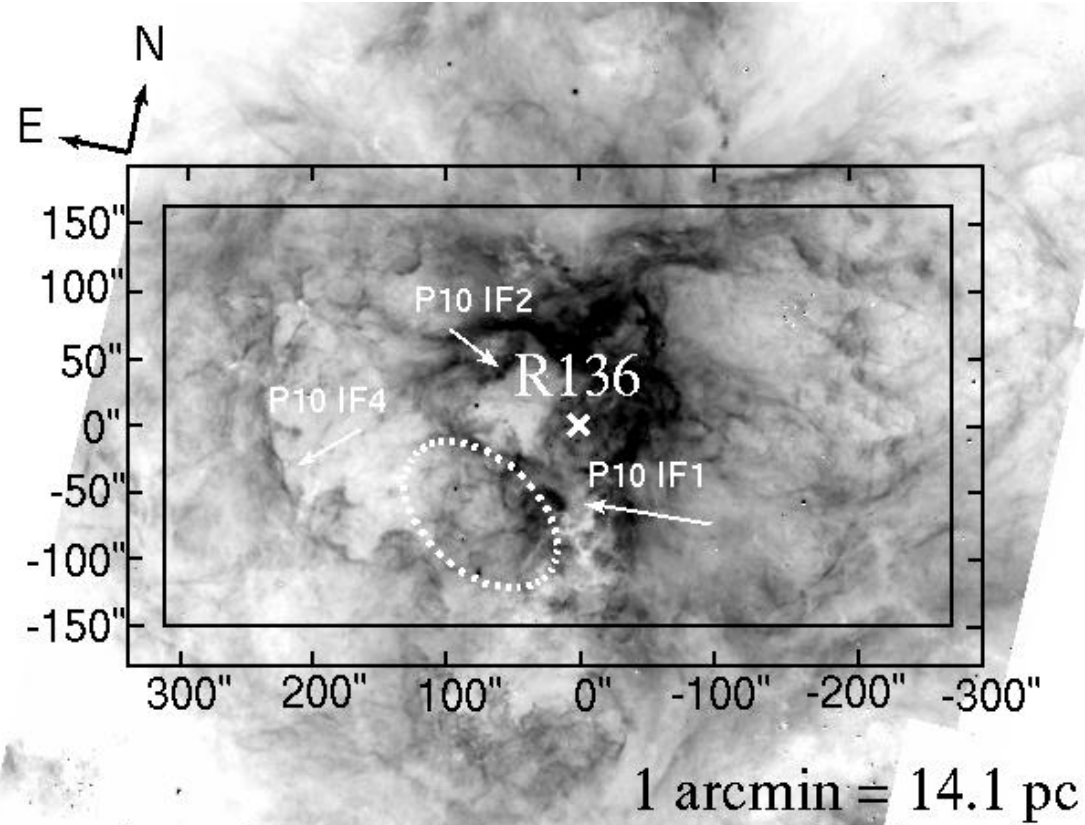

Figure 1- An Hα image taken with SOAR narrow band imaging described in Paper I. A large rectangle identifies the area covered by our survey in Paper I. The position of R 136 is indicated with a cross. The IF's from Paper I that are mentioned in this paper are also labeled, with the notation "P10" indicating Paper I.

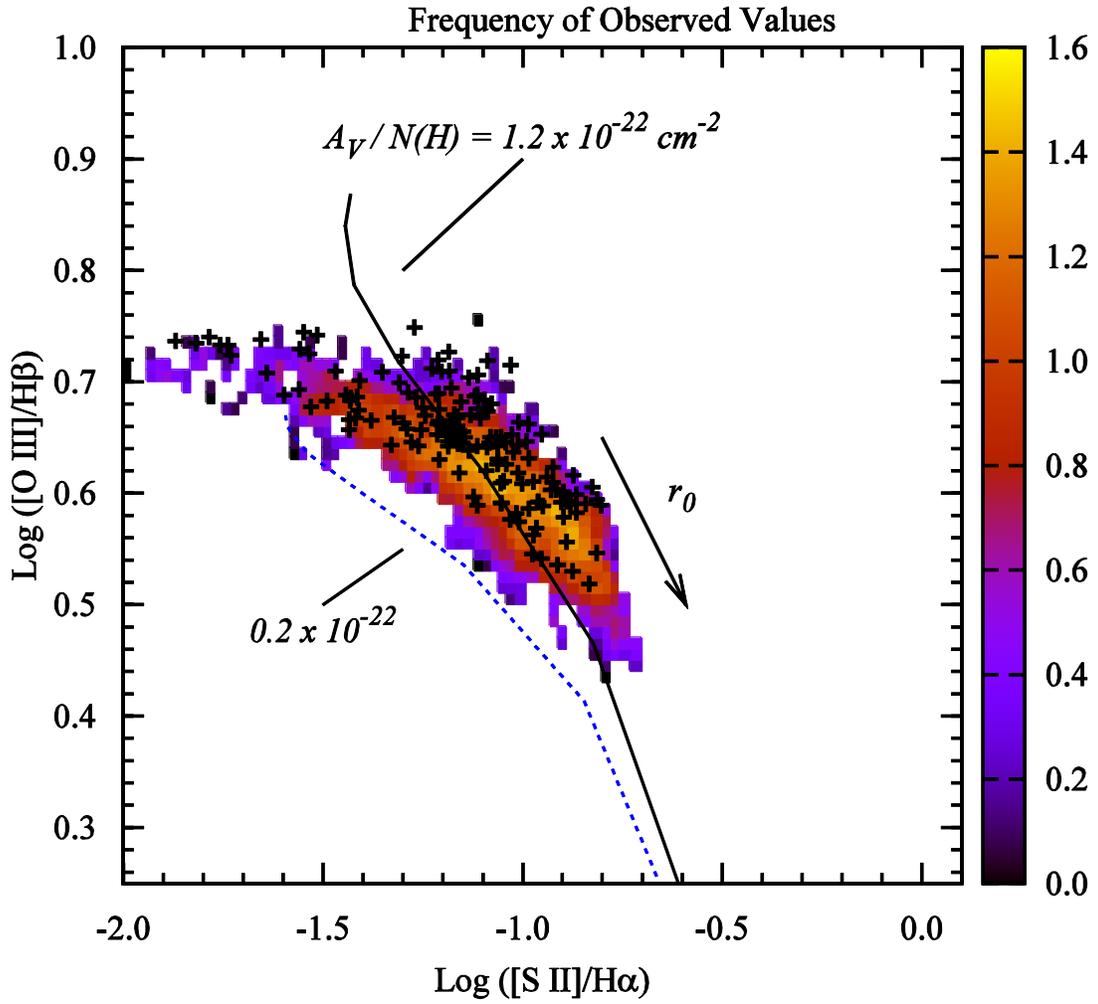

Figure 2 - Plot of the predicted [S II]/Hα vs [O III]/Hβ diagnostic line ratios for models with two dust abundances, but otherwise based on the parameters of TP05 compared with observations from SOAR (black crosses) and Blanco spectra. The Blanco data have been binned to show their distribution. Lines represent photoionization models with $n_e = 200$ cm$^{-3}$. The arrow indicates how the model radius changes.

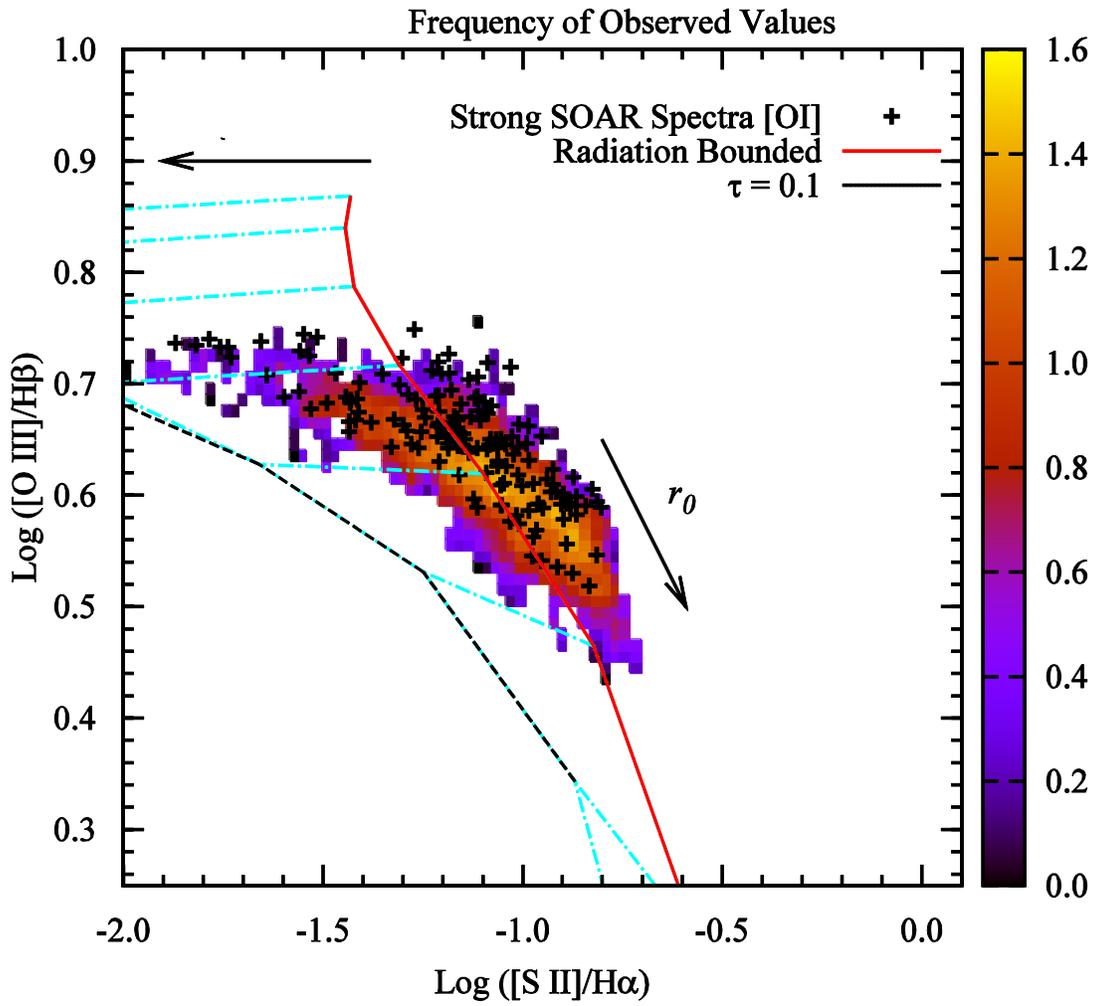

Figure 3 – Similar to Figure 2 with the dust abundance fixed at $A_V/N(H) = 1.2 \times 10^{-22}$ cm$^2$, but with a variable optical depth to ionizing radiation $\tau_{912}$.

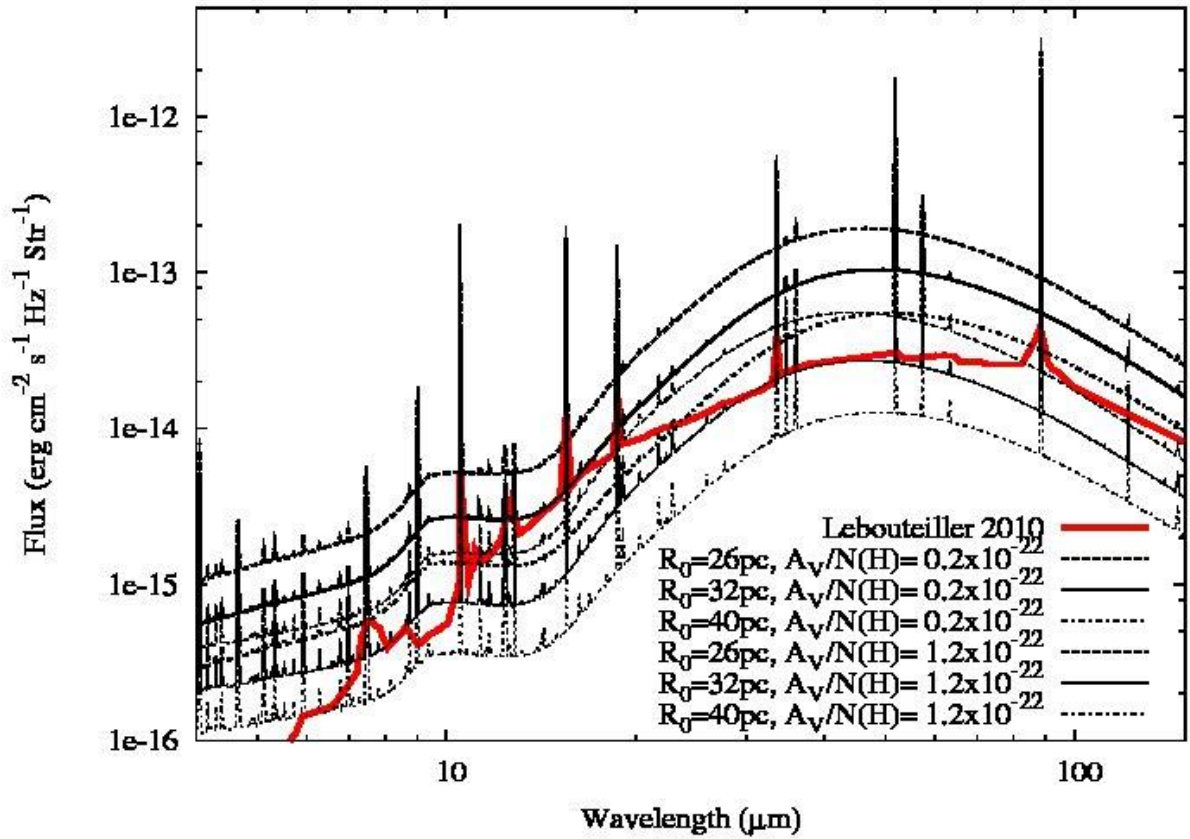

Figure 4 – The observed IR spectrum (thick, red) from 1 to 150 μm of 30 Doradus over a 60 pc square aperture centered on the ionizing cluster NGC 2070. Models with thin lines show predictions using the low dust abundance assumed in the paper, while the medium width lines show predictions for a standard $A_V/N(H)$. The model radii used are chosen to reproduce the observed optical line ratios in that region.

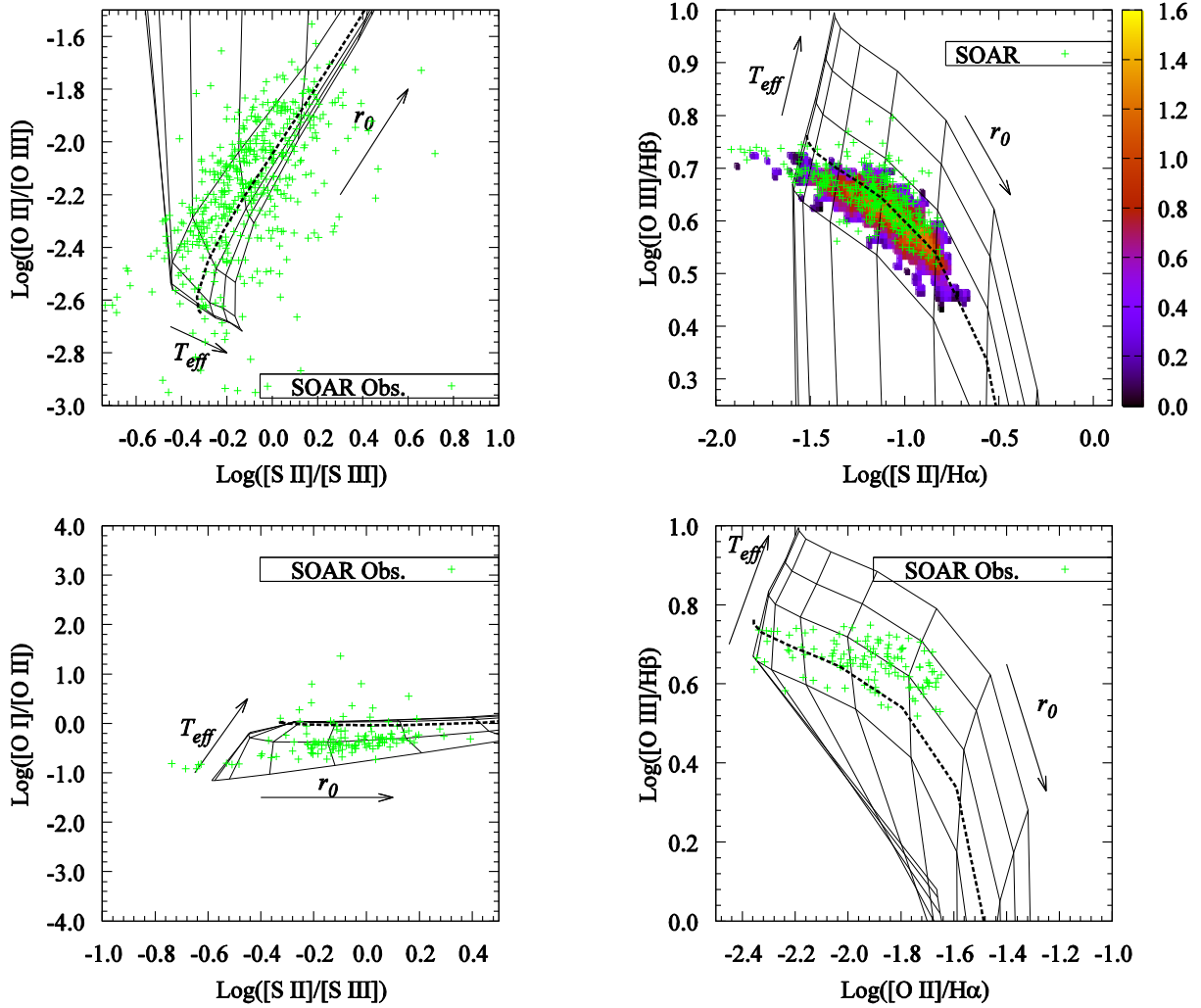

Figure 5 - Plots of diagnostic line ratios with observations from SOAR and Blanco spectra. Lines represent photoionization models with $n_e = 200$ cm$^{-3}$ and the abundances from TP05. Arrows indicate the effect on the line ratios of increasing the modeled parameters $T_{eff}$ and $r_0$. $T_{eff}$ was varied from 36,000K to 42,000K in 2,000K steps. The initial radius $r_0$ was varied from 4 to 158 linear pc in approximately 0.2dex increments. The dashed black line marks 38,500K. From top left to bottom right: (a) [S II]/[S III] vs. [O II]/[O III]; (b) [S II]/Hα vs [O III]/Hβ; (c) [S II]/[S III] vs. [O II]/[O I]; (d) [O II]/Hα vs [O III]/Hβ.

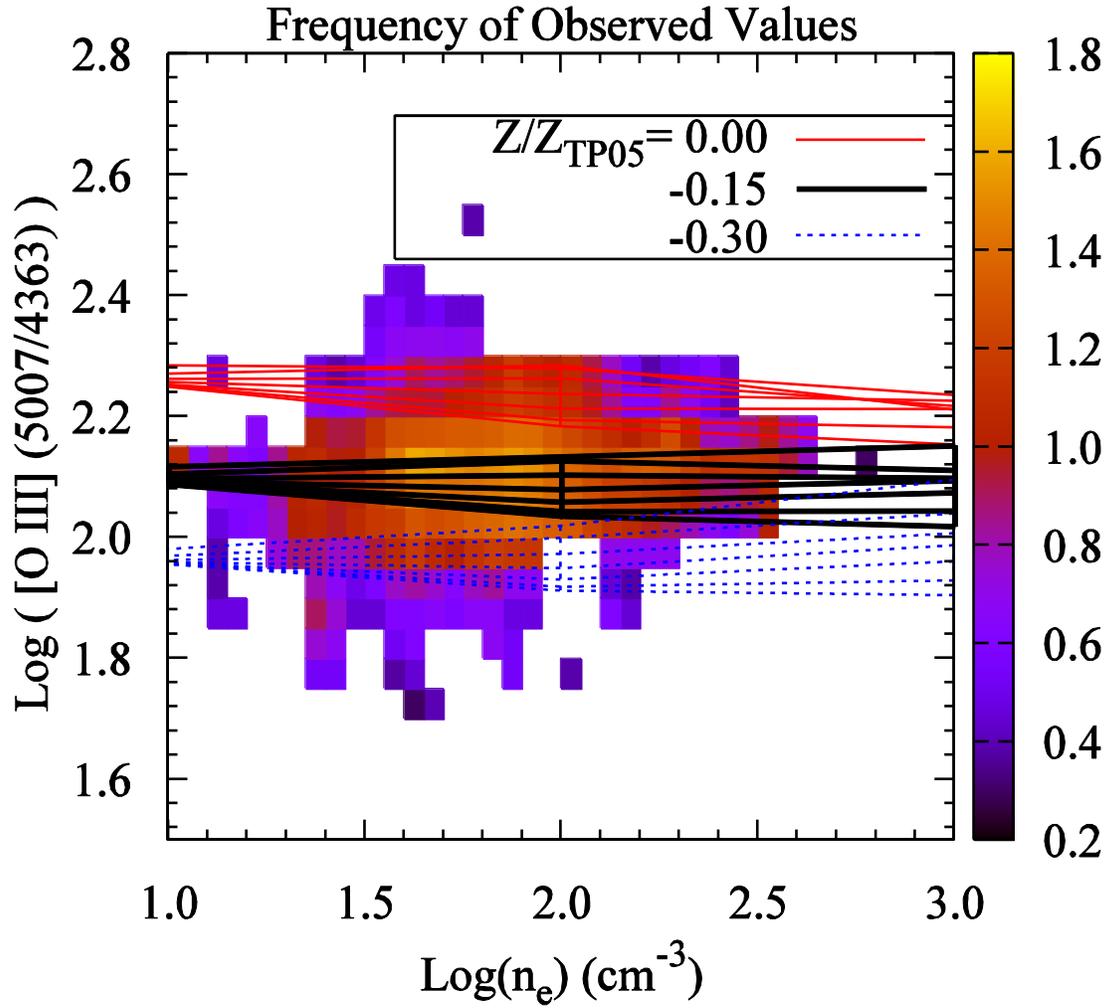

Figure 6 - Predicted and observed [O III] λ5007/λ4363 ratios for models with TP05 abundances for models with $10^1 \leq n(H) \leq 10^3$ cm$^{-3}$. The three grids of models are for the three different sets of metal abundances indicated in the legend. Within each grid, the different curves represent different values of the initial radius $r_0$, varied from 4 (lowest curve) to 158 (highest curve) linear pc in approximately 0.2 dex increments.

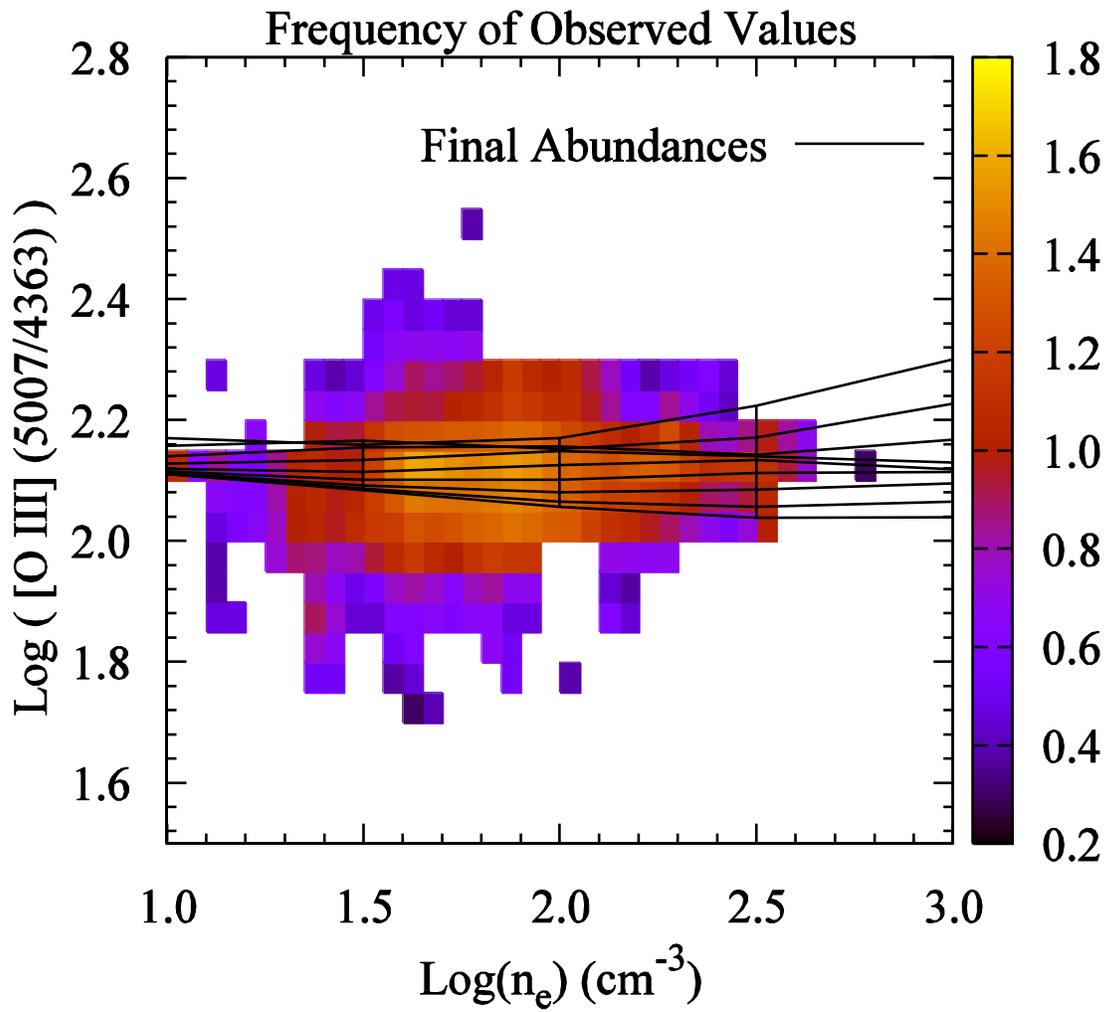

Figure 7 – The [O III] λ5007/λ4363, similar to Figure 5, but with a model grid computed using our final adopted abundances.

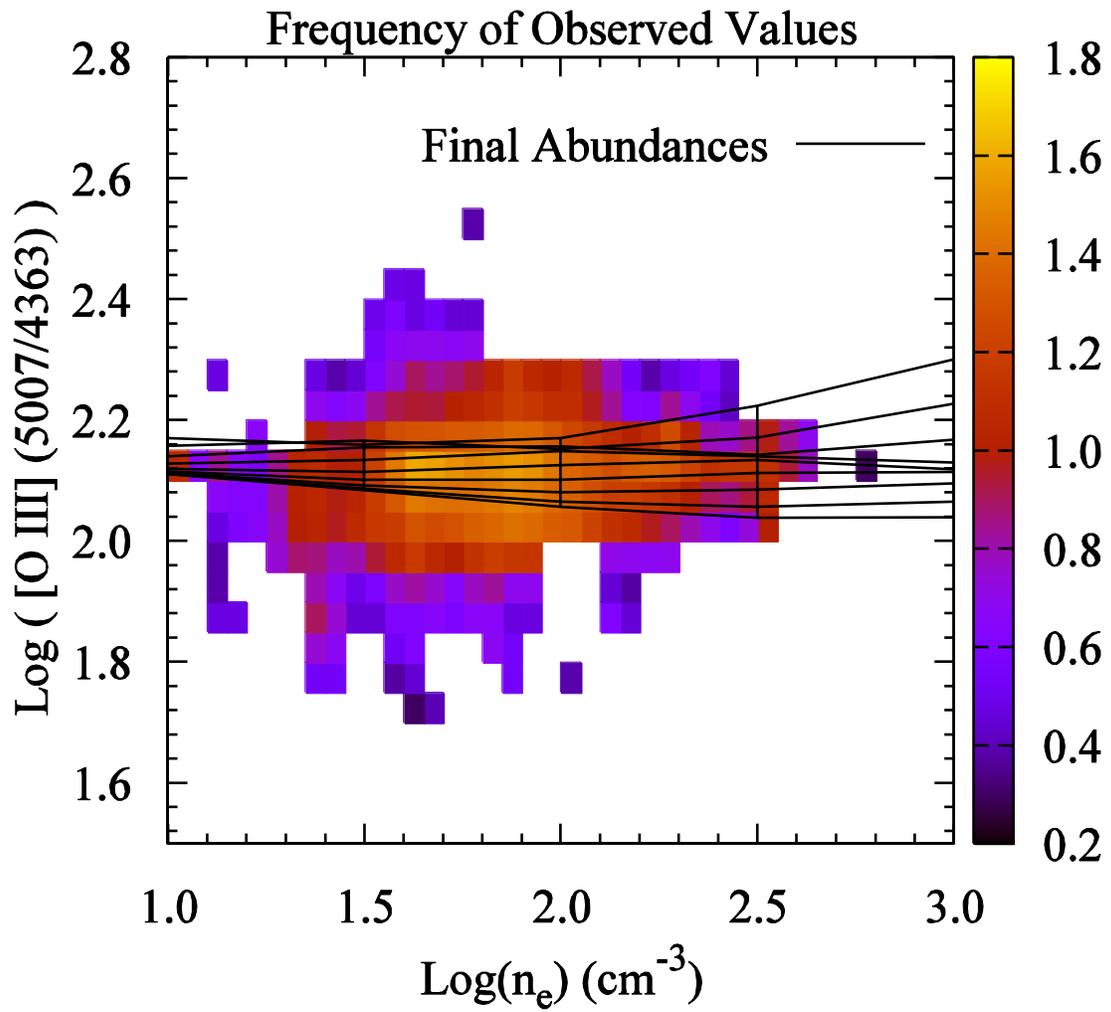

Figure 8 – The same line-ratio diagrams shown in Figure 5, but with model grids using our adopted abundances.

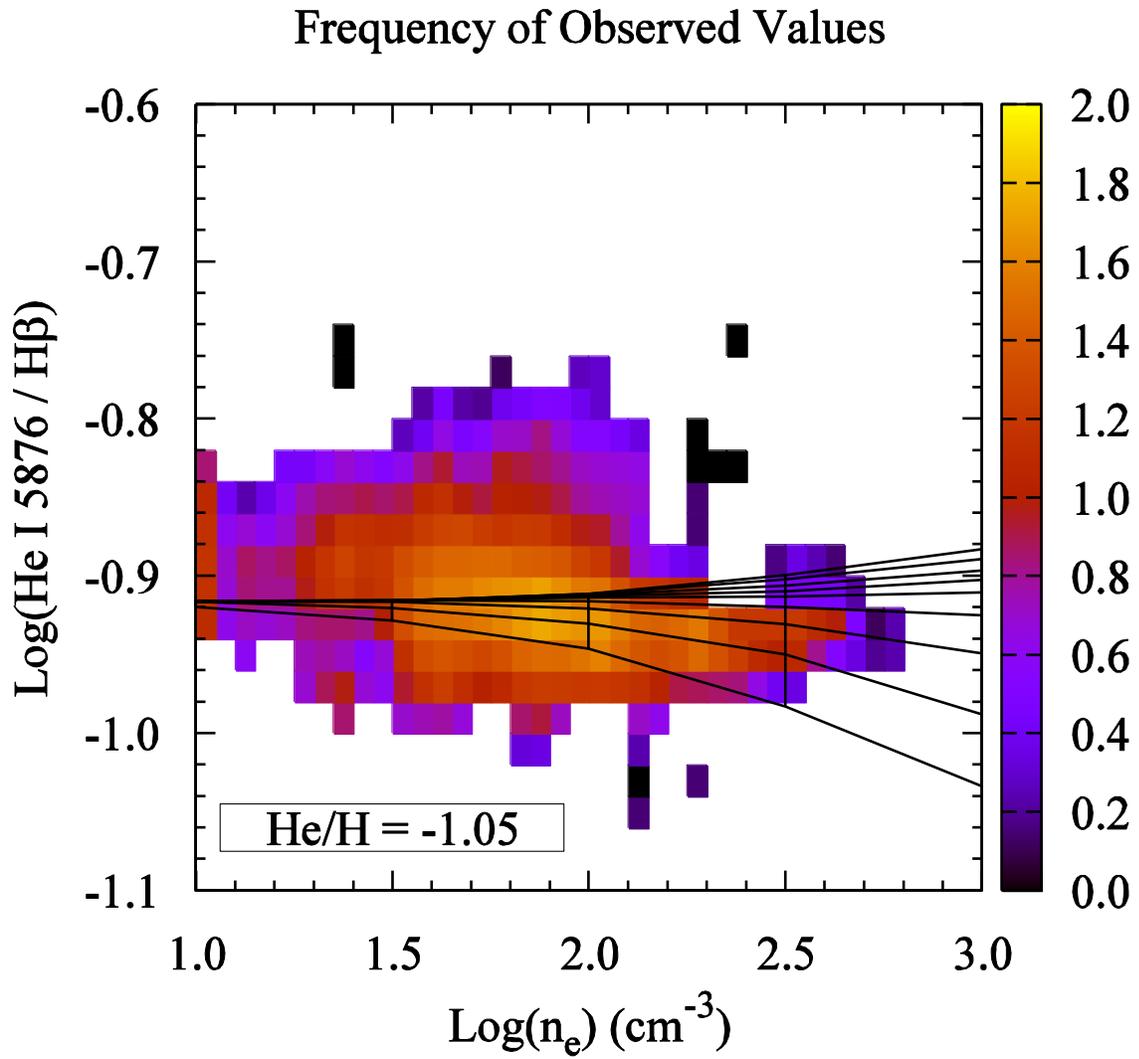

Figure 9 - Predicted and observed intensity ratio of He I λ5876 / Hβ. The different curves represent different values of the initial radius $r_0$, varied from 4 (highest curve) to 158 (lowest curve) linear pc in approximately 0.2 dex increments.

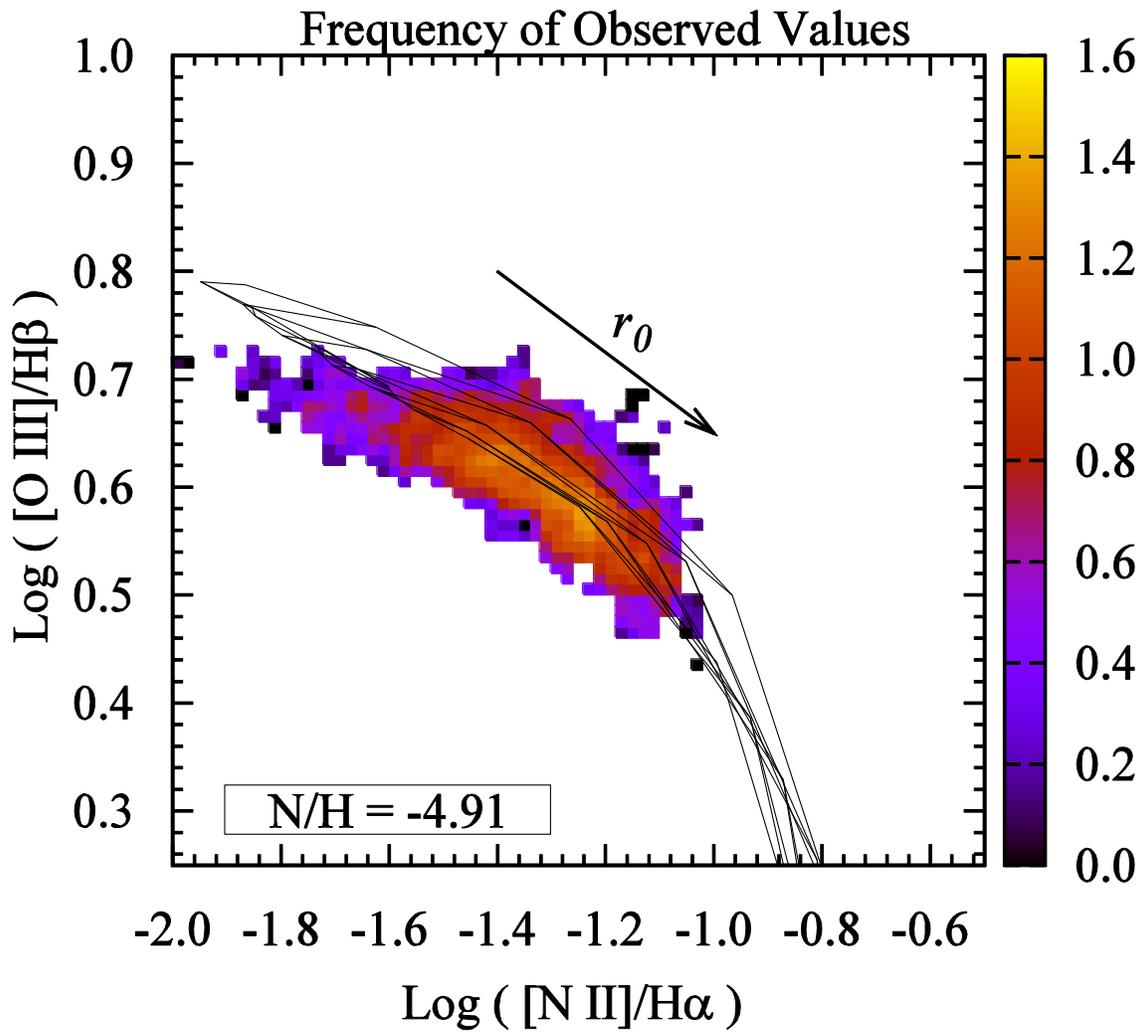

Figure 10 - Predicted and observed intensity ratio of the commonly used [N II] λ6584/Hα vs. [O III]/Hβ diagnostic diagram.

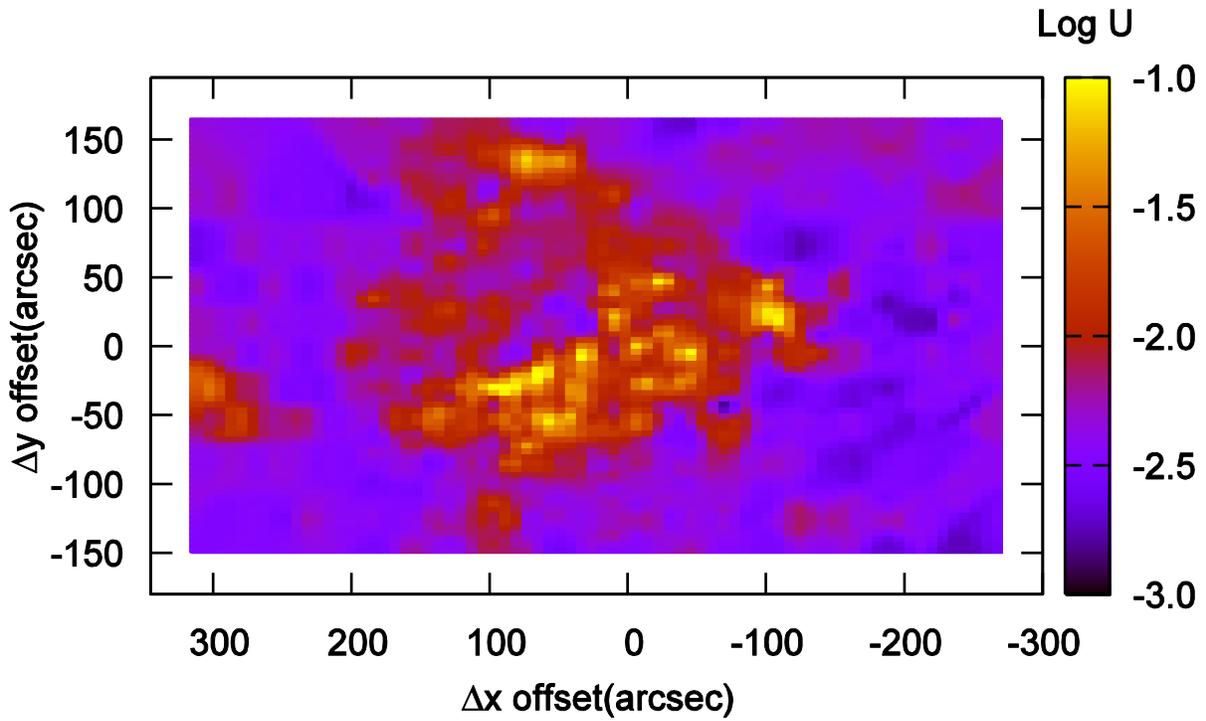

Figure 11 - An interpolated map of the dimensionless ionization parameter *U*, derived from fitting models to our Blanco spectra. The region mapped is the same as that outlined in Figure 1.

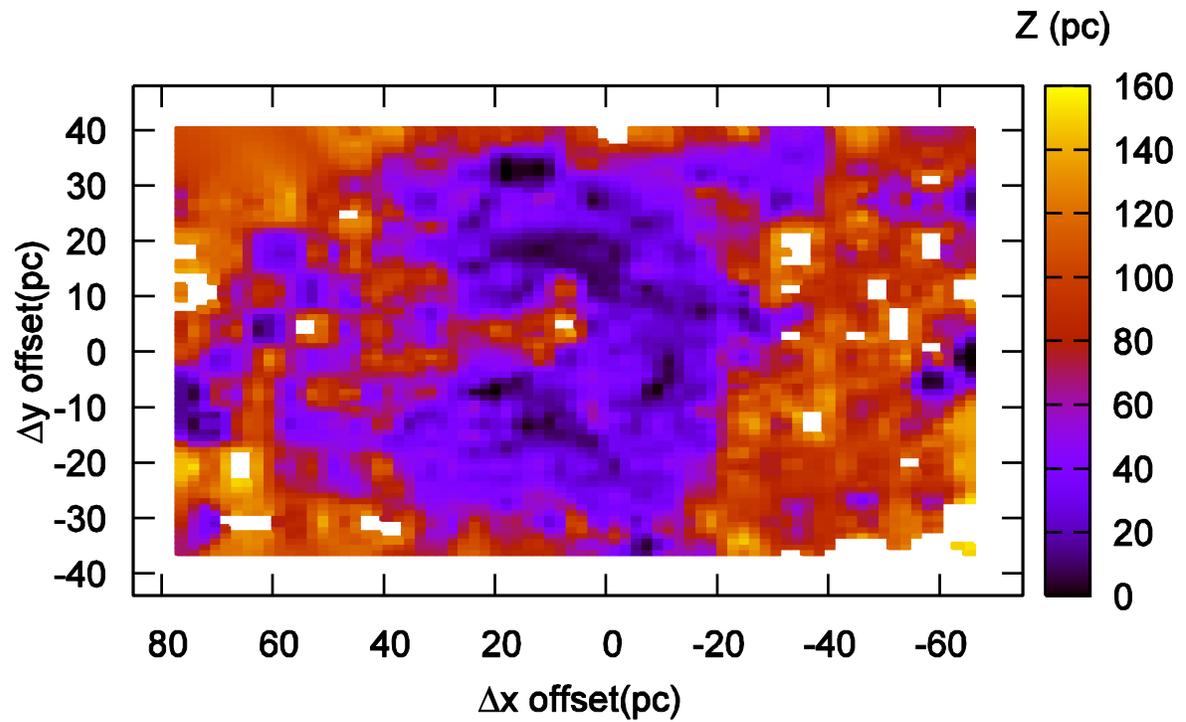

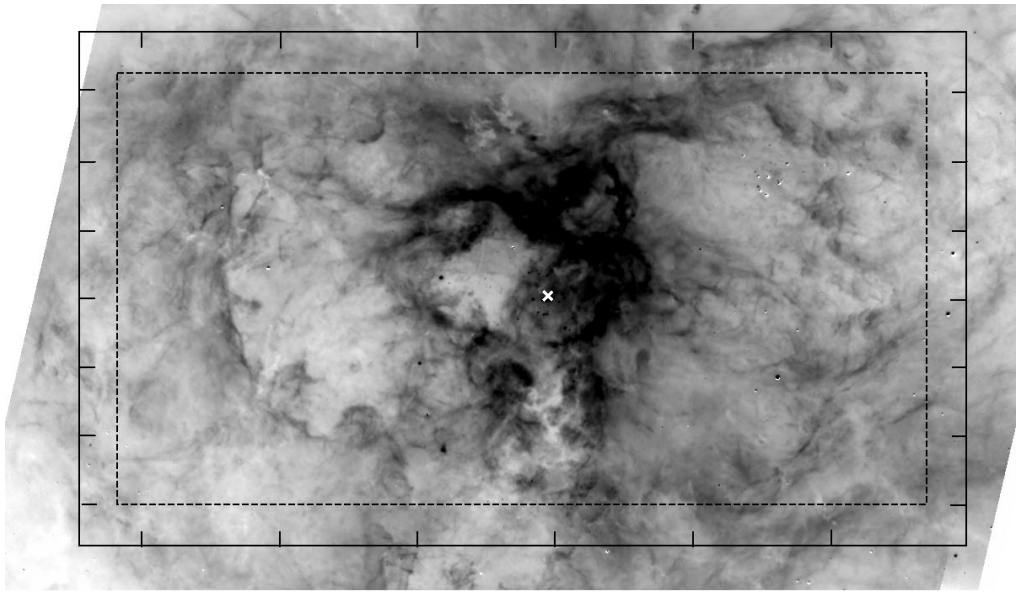

Figure 12 – Map of the line-of-sight position |z|, defined in Eq. 5. By definition, R136 has $z = 0$.

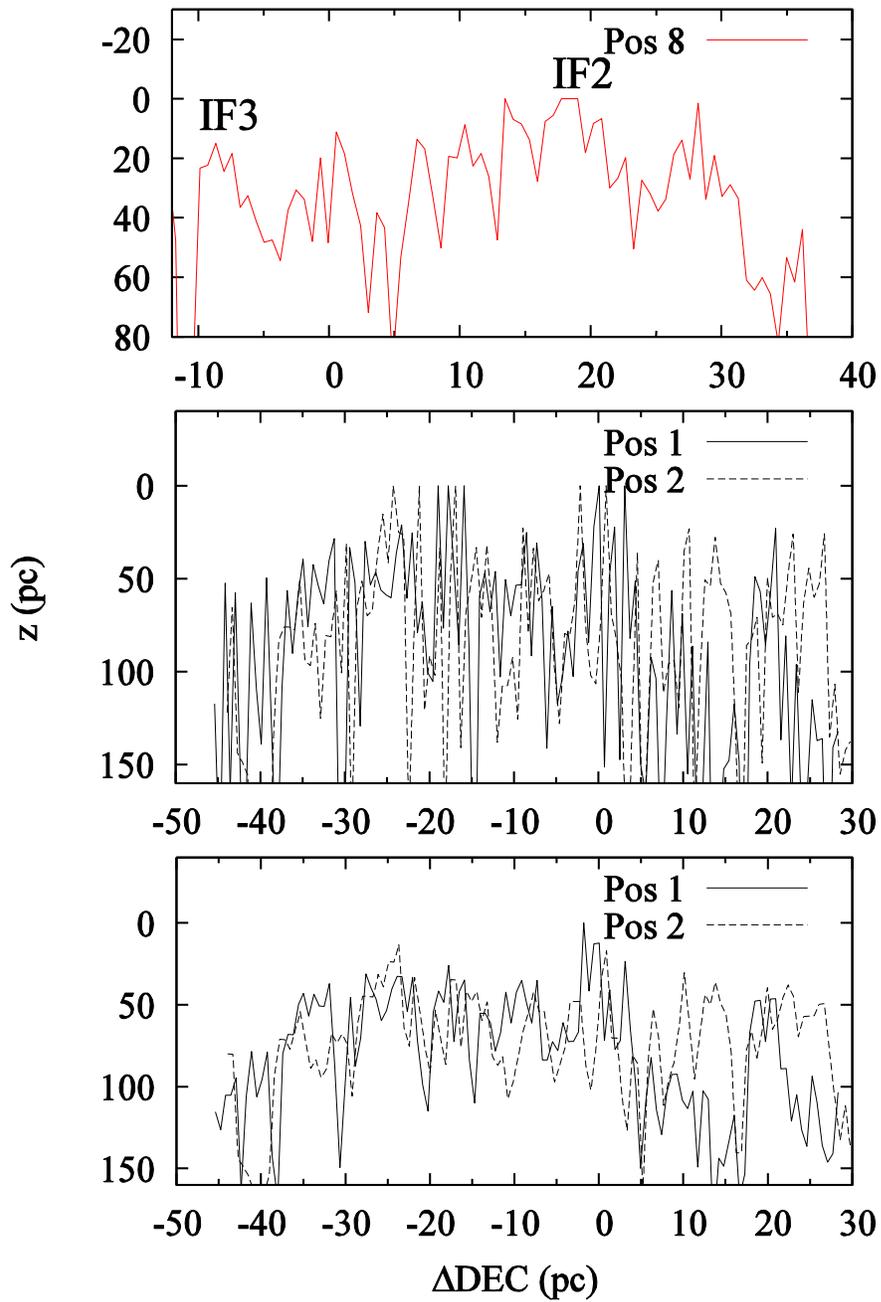

Figure 13 – (top) Profile of the line-of-sight position |z| measured along slit position 8. The x-axis shows the offset in declination from R136. By definition the height of NGC 2070 and R136 is $z = 0$. (middle) Profile of |z| for slit positions 1 and 2, using the best fitting model; (bottom) |z| calculated from Equation 5 assuming a smooth density distribution.

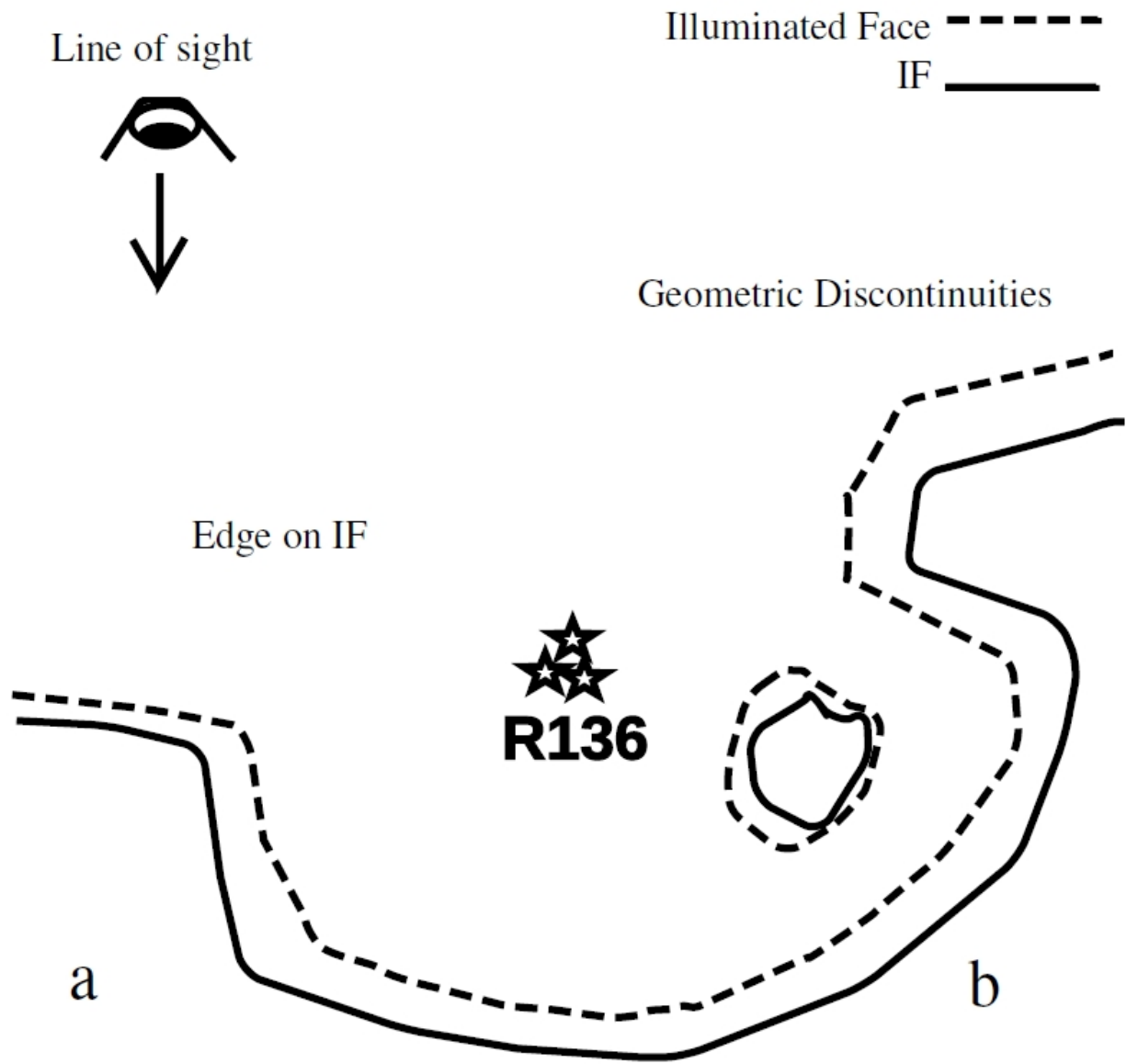

Figure 14 - A cartoon of possible geometries in 30 Doradus consistent with the changes in $U$ across the nebula. The region labeled "a" is a continuous ionization front of finite height, facing the ionizing cluster but seen edge-on. Region "b" represents possible geometries that would be seen as discontinuities in modeled $r_0$ and $|z|$.

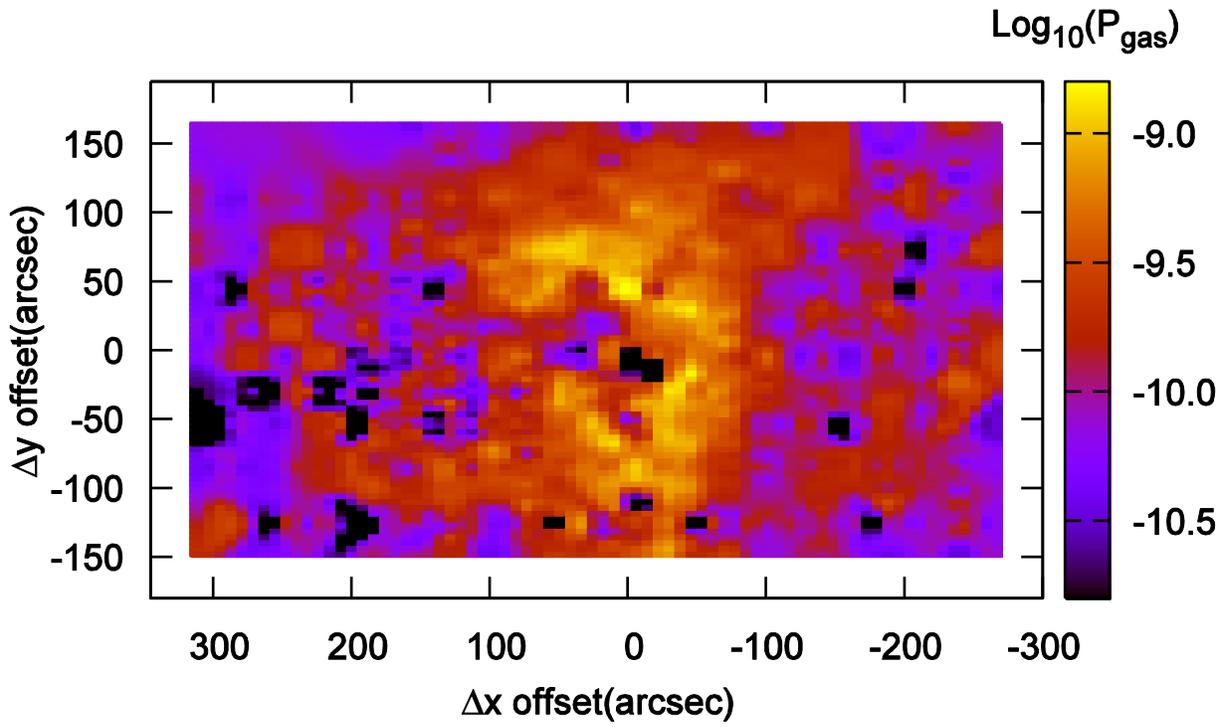

Figure 15 - Observed thermal gas pressure, in units dyne cm$^{-2}$, interpolated from the Blanco spectra.

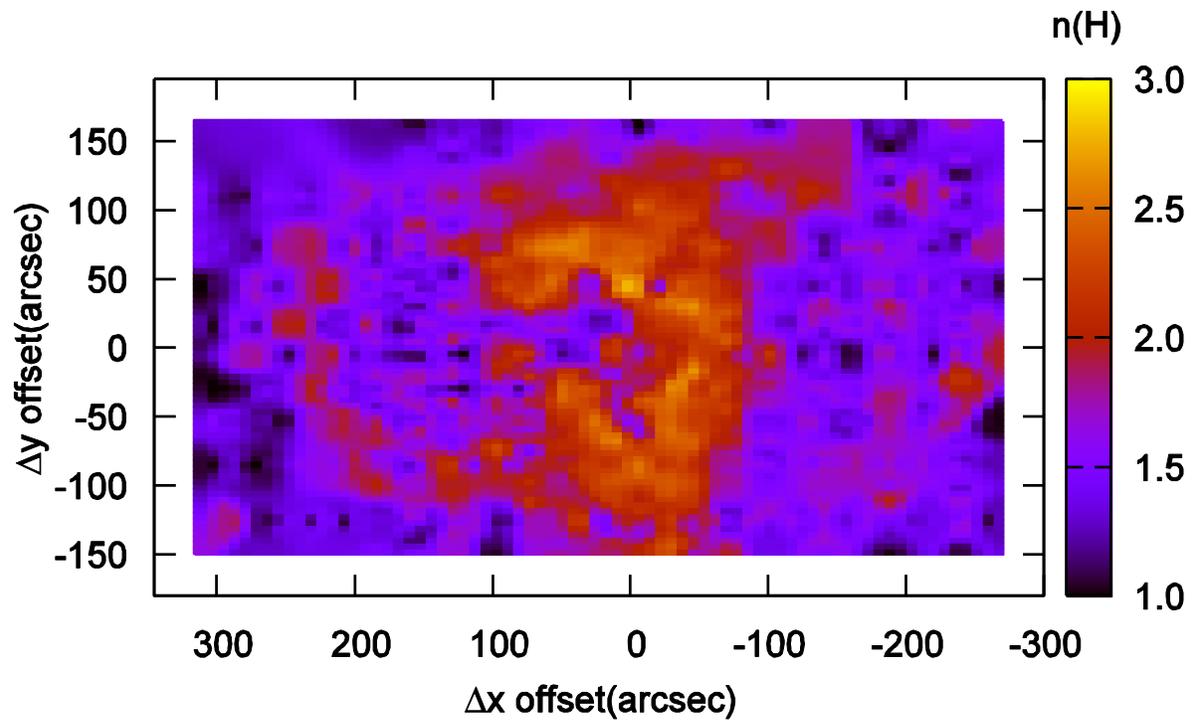
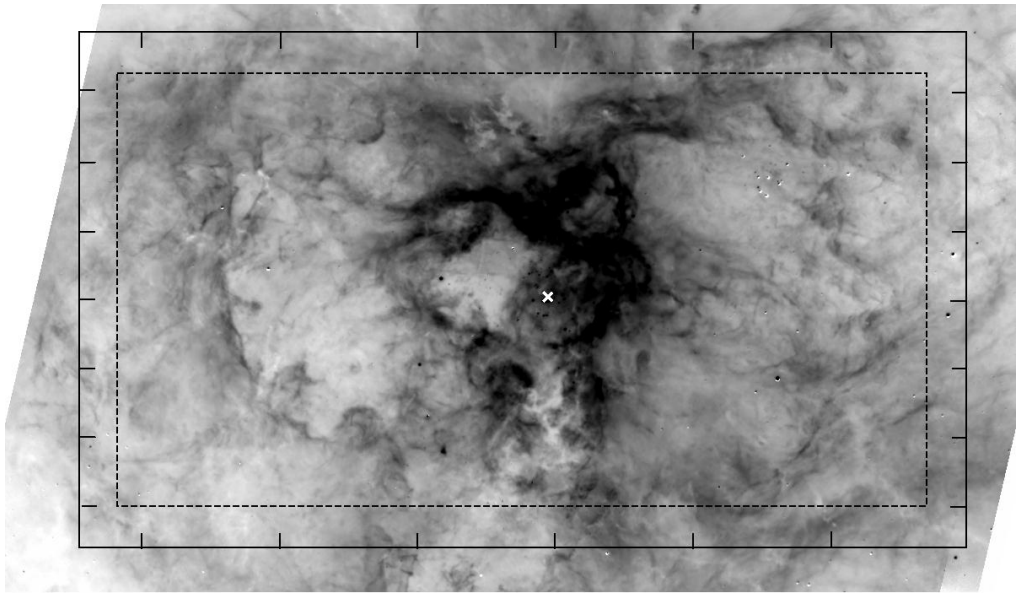

Figure 16 - Map of $n_e$ made with measurements of the [S II] 6717/6731 doublet from data in Paper I.

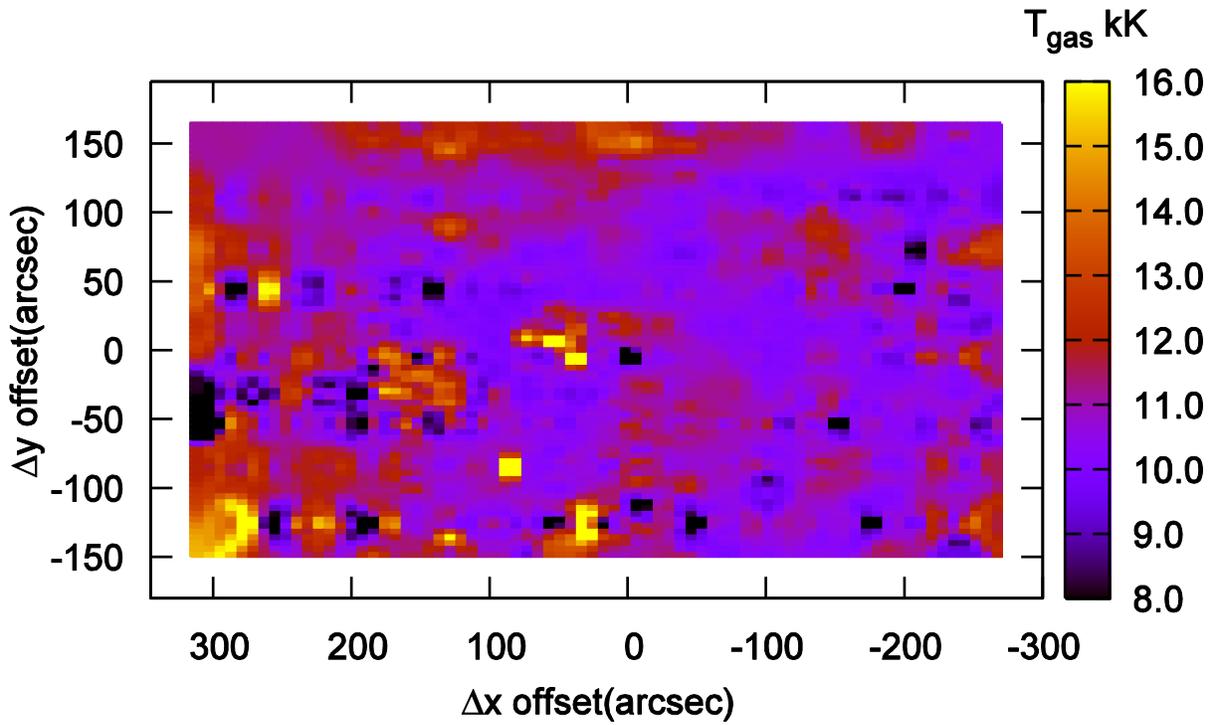

Figure 17 – Map of $T_e$ from the [O III] diagnostic [O III] 5007/4363 using our data Paper I. Generally the nebula is isothermal to ±1000s degrees, with the coolest gas being of the highest density.

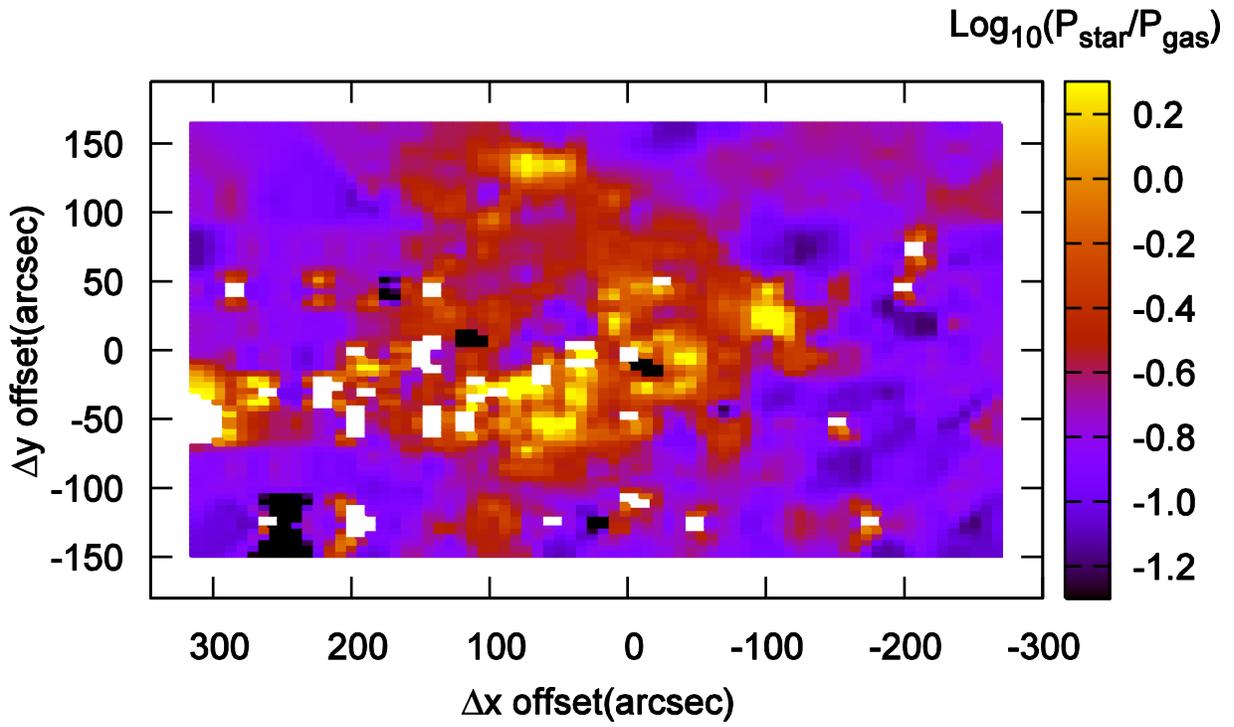

Figure 18 – The ratio of pressure from integrated star light to gas pressure.

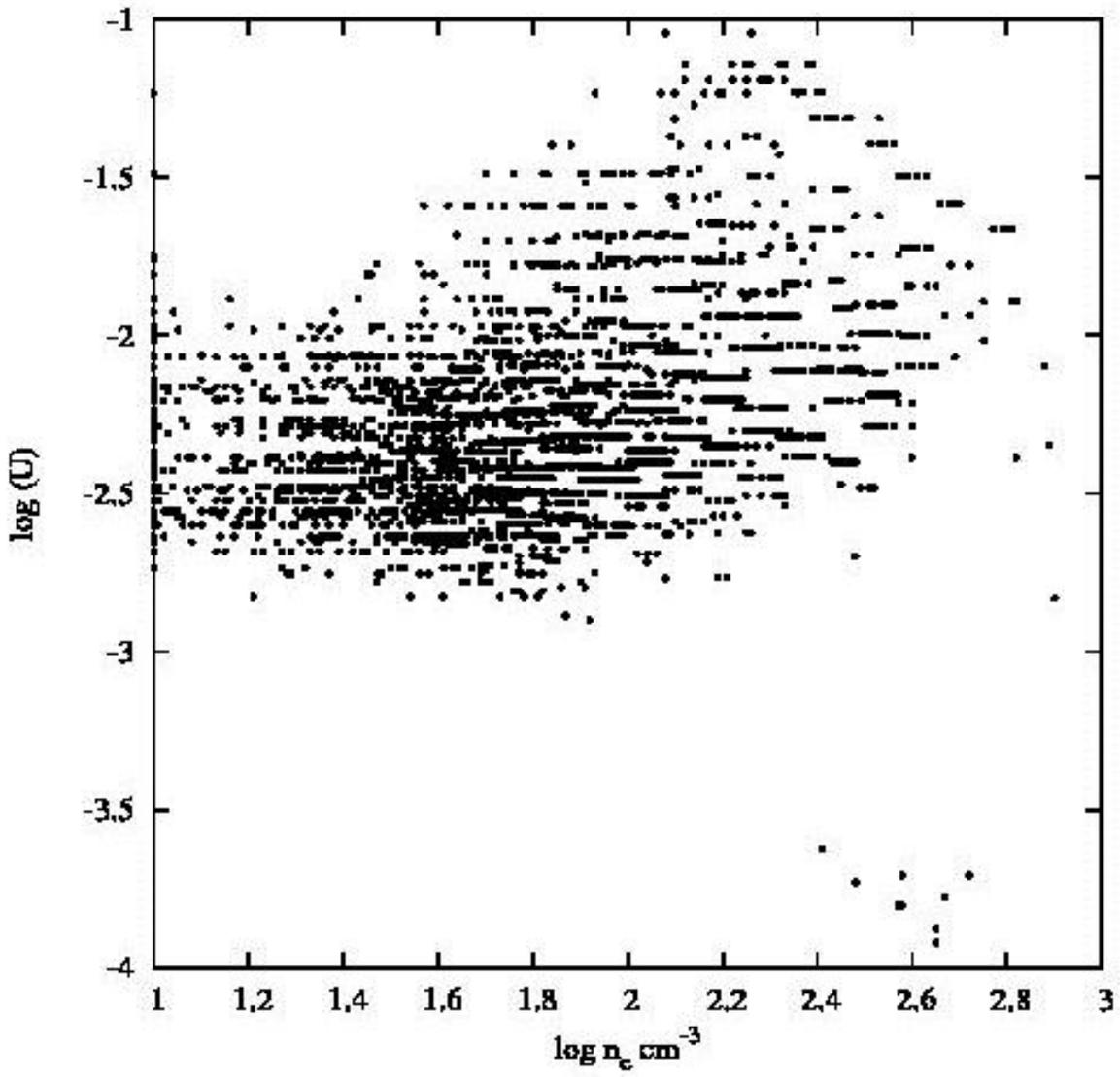

Figure 19 - Observed gas density vs. modeled ionization parameter *U*.

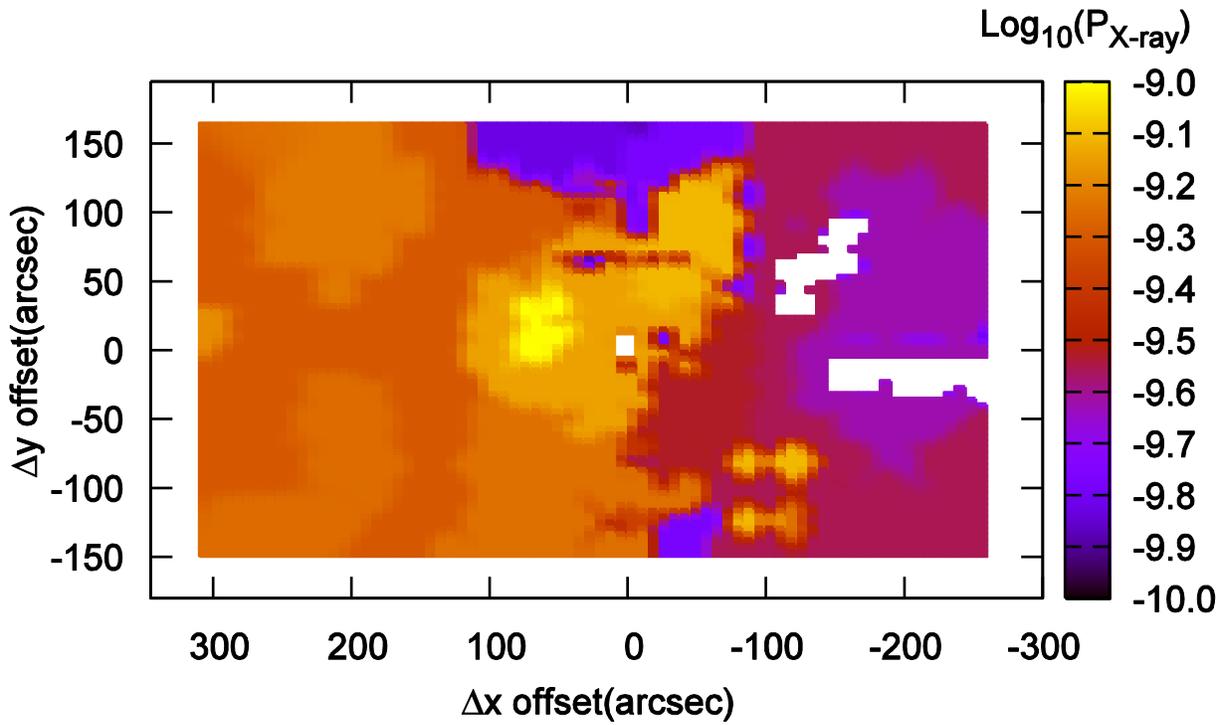

Figure 20 – $P_{\text{X-ray}}$, in units of dyne cm$^{-2}$, from the regions of diffuse x-ray emission described in Townsley et al. (2006). The pressure was calculated using the reported $T_{\text{x-ray}}$, surface brightness and area.

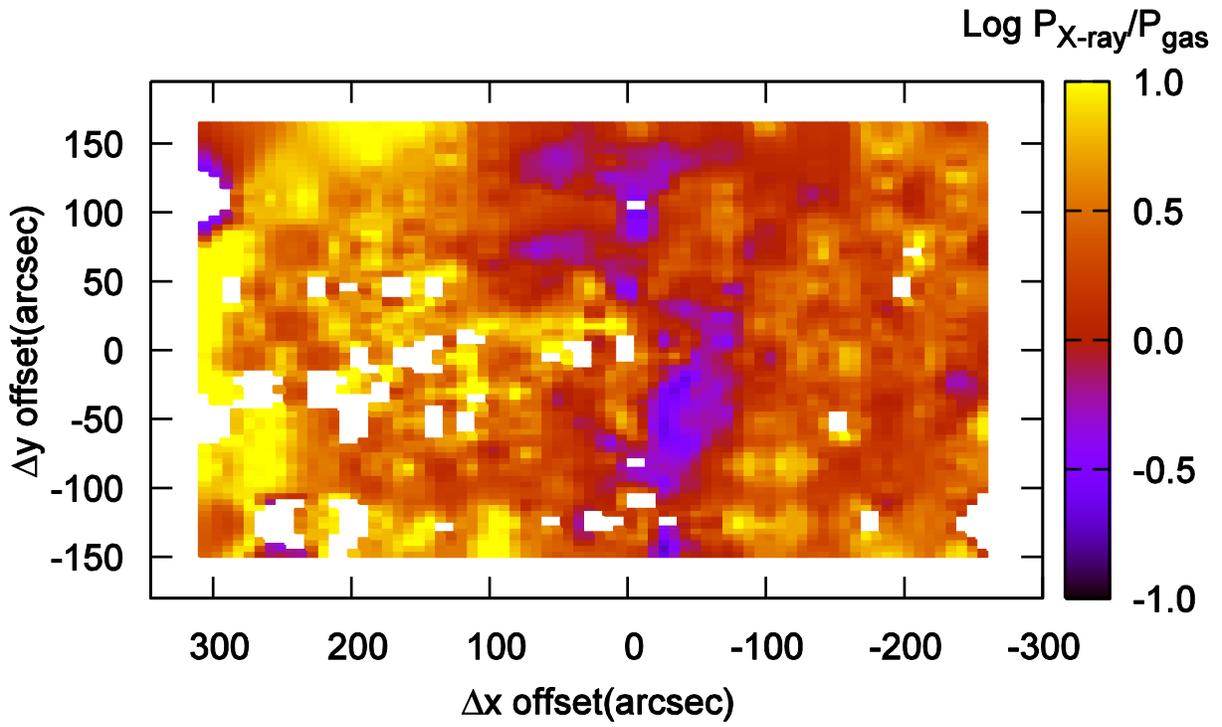

Figure 21 – The ratio of $P_{\text{x-ray}}$ to $P_{\text{gas}}$.

## Appendix A. Dust within the 30 Dor Complex

Our simulations of 30Dor for the first time include a fully self-consistent treatment of gas-grain-photon interactions. Grains are heated by gas-grain collisions, but most importantly, by absorption of the radiation field striking the particles. Photoelectric heating is important for gas in the H II region (Baldwin et al. 1991; hereafter BFM) and in the PDR (Tielens 2005). The radiation field includes the stellar spectral energy distribution, attenuated by gas and dust opacities within the nebula, and lines and continua emitted within the nebula. BFM give an early overview of our treatment, and van Hoof et al. (2004) describe more recent revisions. These include resolved grain-size distributions and improved grain-charging based on Weingartner & Draine (2001b). Other details are given in Abel et al. (2008).

The effects of grains upon the ionized gas depend on the ionization parameter, with highly ionized regions most strongly affected (Bottorff et al. 1998). This is basically because the grain opacity does not depend on the ionization of the gas, while the gas opacity decreases as the level of ionization increases, since there are fewer atoms to absorb radiation. The effect is that grains have their greatest influence upon high-ionization nebulae (Bottorf et al. 1998).

The grain-size distribution also affects the amount of photoelectric heating of the gas. Small grains have the largest ratio of area to volume so contribute a disproportionate amount of extinction relative to their mass. They are the grains most efficient at heating the gas for this reason (Tielens 2005). We use a distribution of grain sizes with a relatively large population of small grains to reproduce the observed LMC extinction curve (Weingartner & Draine 2001a). As a result our grains are especially efficient at heating the gas.

The dust-to-gas ratio in the diffuse ISM of the LMC, $A_V/N(H) = 1.2 \times 10^{-22}$, has been measured in a large number of studies. Tests show that photoionization models that use this ratio, with the LMC grain-size distribution, produce emission-line ratios that have a large upward "hook" at the high ionization, left hand end, of the emission-line ratio diagram (Figure 2). This hook is not observed, and, in fact, brings the models into disagreement with the observations. The origin of the predicted hook is that grain photoelectric heating raises the temperature of the gas in this region of the diagram, causing the forbidden lines to become too strong relative to the recombination lines. This conflict with observations suggests that either the current grain heating theory overestimates the heating, or the dust abundance is smaller than measured in the LMC diffuse ISM.

It is doubtful that the current grain theory overestimates the gas-grain photoelectric heating. Allers et al. (2005) studied molecular hydrogen emission lines across the Orion "Bar", a nearby ionization front seen nearly edge on. This region is largely heated by grain-electron photoejection. Allers et al found that the gas is observed to be warmer than predicted with the Weingartner & Draine (2001a) theory, and suggested that this theory may need to be revisited, with the aim of *increasing* the heating. That would make our disagreement worse. Shaw et al. (2009) and Pellegrini et al. (2009a) used our spectral simulation code to confirm this result, and proposed that enhanced cosmic rays may account for the warmer-than-expected temperatures. Lowering the heating efficiency, to fix our 30 Dor simulations, would make the problems found by Allers et al in Orion even worse.

The only other alternative to changing grain photoelectric heating theory is to lower the dust to gas ratio in the ionized region. We did so, finding that values below $0.2\times10^{-21}$ cm$^{-2}$ are needed to remove the "hook" in Figure 2. We assume this dust-to-gas ratio in the body of this paper.

We do not address the origin of this low dust to gas ratio. The low abundance could be produced in one of two ways – the grains may have been destroyed in the ionized gas or pushed beyond it.

We determine the grain temperature for each size and material type, including quantum temperature pulsing. The grains remain well below their sublimation temperatures throughout the cloud, so all sizes should survive. However they may have been destroyed in transient events such as passing shock waves (although shocks *are not* important in powering the emission lines). If they are destroyed their constituents would be released into the gas phase. This would have significant effects on the optical spectrum, mainly producing strong [Fe II] and [Ca II] emission (Shields 1975; Kingdon, Ferland & Feibelman 1995). This is not present, suggesting that the majority of the grains have not been destroyed. Paradis et al. (2009) have shown evidence suggesting shattering of large dust grains into smaller ones in the 30 Dor region, which could account for the small *R*, but would not alter the total amount of dust present.

A second possibility is that the grains have been pushed out of the ionized gas (Ferland 2001). We determine the gas-grain relative drift velocity as part of our simulations (BFM). The speeds are substantially smaller than the gas sound speed because of efficient gas-grain Coulomb coupling due to grain charging. Gas-grain separation is not important today, and could only have occurred before ionizing radiation was present.

Neither destruction nor separation appears to be viable so we leave the low deduced dust to gas ratio as a conundrum.